\documentclass[aos,preprint]{imsart}
\usepackage{amsmath,amssymb,amsthm,mathtools}
\usepackage[authoryear]{natbib}
\usepackage{booktabs}
\usepackage{graphicx}
\usepackage[colorlinks=true,citecolor=blue,urlcolor=blue,linkcolor=blue]{hyperref}

\startlocaldefs
\numberwithin{equation}{section}
\theoremstyle{plain}
\newtheorem{theorem}{Theorem}[section]
\newtheorem{proposition}[theorem]{Proposition}

\newtheorem{corollary}[theorem]{Corollary}
\theoremstyle{definition}
\newtheorem{assumption}[theorem]{Assumption}

\newtheorem{example}[theorem]{Example}
\newcommand{\R}{\mathbb R}
\newcommand{\C}{\mathbb C}
\newcommand{\E}{\mathbb E}
\newcommand{\Prob}{\mathbb P}
\newcommand{\norm}[1]{\left\lVert#1\right\rVert}
\newcommand{\ip}[2]{\left\langle#1,#2\right\rangle}

\newcommand{\argmin}{\operatorname*{arg\,min}}
\newcommand{\tr}{\operatorname{tr}}

\newcommand{\bN}{\bar N}
\newcommand{\bm}{\bar m}

\newcommand{\NumLamThreeInv}{81}
\newcommand{\NumIdifFit}{0.048}

\newcommand{\NumGamma}{0.65}
\newcommand{\NumRcritOne}{3{,}809}
\newcommand{\NumRcritTwo}{239}

\newcommand{\NumFrEndAfter}{31}

\newcommand{\NumFrLatest}{10}

\newcommand{\NumBdReps}{8}
\newcommand{\NumBdLamMeas}{1.71 (0.05)}
\newcommand{\NumBdLamPred}{1.63}
\newcommand{\NumBdLamCurv}{0.074}
\newcommand{\NumBdSigLo}{0.025}

\newcommand{\NumCondLminRange}{$[0.0008,0.0025]$}
\newcommand{\NumCondRange}{$[6.9,43.3]\times10^3$}
\newcommand{\NumCondAtA}{0.123}

\newcommand{\NumPopRatiosText}{1.7--7.1}

\newcommand{\NumValGamma}{0.65}
\newcommand{\NumValCondN}{1.1\times10^5}
\newcommand{\NumValAtA}{0.043}

\newcommand{\NumValBpred}{(0.01, $-1.63$, $-3.08$, 1.29, 1.81, $-3.00$)}
\newcommand{\NumValPredA}{46\%}
\newcommand{\NumValPredB}{17\%}
\newcommand{\NumValPredC}{1\%}
\newcommand{\NumValNA}{120}

\newcommand{\NumValCovJA}{36\%}

\newcommand{\NumValProjJA}{1.14}
\newcommand{\NumValProjJAse}{0.06}

\newcommand{\NumValCovCA}{88\%}
\newcommand{\NumValCovCACI}{81--93\%}
\newcommand{\NumValProjCA}{0.26}

\newcommand{\NumValNB}{80}

\newcommand{\NumValCovJB}{8\%}

\newcommand{\NumValProjJB}{1.16}
\newcommand{\NumValProjJBse}{0.04}

\newcommand{\NumValCovCB}{88\%}
\newcommand{\NumValCovCBCI}{78--94\%}
\newcommand{\NumValProjCB}{0.22}

\newcommand{\NumValNC}{30}

\newcommand{\NumValCovJC}{0\%}

\newcommand{\NumValProjJC}{1.13}
\newcommand{\NumValProjJCse}{0.04}

\newcommand{\NumValCovCC}{97\%}
\newcommand{\NumValCovCCCI}{83--100\%}
\newcommand{\NumValProjCC}{0.16}

\newcommand{\NumValOracleNclose}{26}
\newcommand{\NumValOracleMax}{0.70}
\newcommand{\NumValFineMax}{$10^{-3}$}

\newcommand{\NumValStartMed}{18}
\newcommand{\NumValStartMax}{117}

\newcommand{\NumPopRefSmall}{10.5\%}

\newcommand{\NumRkChecks}{2{,}016}

\newcommand{\NumRkMaxErr}{$7.9\times10^{-10}$}

\newcommand{\NumDecCovLoA}{68}
\newcommand{\NumDecCovHiA}{92\%}
\newcommand{\NumDecSdLoA}{1.20}
\newcommand{\NumDecSdHiA}{1.97}

\newcommand{\NumDecCovLoB}{86}
\newcommand{\NumDecCovHiB}{96\%}

\newcommand{\NumDecCovHiC}{97\%}

\newcommand{\NumSnPairs}{30}
\newcommand{\NumSnR}{1024}

\newcommand{\NumSnSdseCoordBy}{0.70--1.27, 0.77--1.10, 0.76--1.09}

\newcommand{\NumSnJoint}{1.19, 1.09, 1.05}
\newcommand{\NumSnCorr}{0.32, 0.11, 0.05}
\newcommand{\NumSnCorrSe}{0.02, 0.02, 0.02}
\newcommand{\NumSnCorrBias}{1.30, 0.23, 0.05}

\newcommand{\NumSnSdse}{0.95, 1.01, 0.99}

\newcommand{\NumBdShortText}{In 68 data sets a channel ended at the $\log\beta$ box $[-6,3]$, near the faces $\beta\to0$ and $\beta\to\infty$ of the macro family, where the channel information degenerates. Refitted with the box $[-9,5]$, 16 joint estimates moved by more than half a standard error to other local minima, and the plug-in correction changed by up to 44 standard errors; the corrected coverage at $R=256$ fell from 106 to 88 of 120 (103 with a face-specific correction). The box restricts the numerical domain without by itself verifying Assumption~\ref{ass:reg}, and the correction is sensitive to how close it lets channels come to these faces.}

\newcommand{\NumOptShortText}{Refits of the 31 data sets with the largest scores or two local minima, with tighter tolerance, an extended valley search and a second start, reduced the largest score from 0.51 to $7.5\times10^{-3}$, moved the estimates by at most 2.26 standard errors and changed the corrected projection coverage at $R=256$ from 106 to 105 of 120, so optimization error of this size does not explain the residual bias or the coordinatewise undercoverage.}
\newcommand{\NumOptDiscText}{Tighter refits of the data sets with the largest profile scores moved the estimates by at most 2.26 standard errors and the projection coverage changed from 43 to 44 of 120 (joint, $R=256$) and from 106 to 105 of 120 (corrected, $R=256$).}
\newcommand{\NumVbFailed}{none}
\newcommand{\NumVbN}{800}

\newcommand{\NumVbKappa}{0.01}

\newcommand{\NumVbCovA}{85\%}
\newcommand{\NumVbCovACI}{81--88\%}
\newcommand{\NumVbCovHA}{90\%}

\newcommand{\NumVbCoordA}{72--95\%}
\newcommand{\NumVbCoordHA}{89--96\%}
\newcommand{\NumVbSdA}{1.09--1.93}
\newcommand{\NumVbSdHA}{1.01--1.27}

\newcommand{\NumVbBiasA}{0.69}

\newcommand{\NumVbCovB}{93\%}
\newcommand{\NumVbCovBCI}{88--96\%}

\newcommand{\NumVbCoordB}{91--97\%}

\newcommand{\NumVbSdB}{0.94--1.15}

\newcommand{\NumVbCovJC}{92\%}
\newcommand{\NumVbCovC}{94\%}
\newcommand{\NumVbCovCCI}{89--96\%}

\newcommand{\NumVbCoordC}{93--96\%}

\newcommand{\NumVbSdC}{0.99--1.08}

\newcommand{\NumVbRegSmoothMax}{0.49}
\newcommand{\NumVbMultiN}{120}
\newcommand{\NumVbMultiFar}{four}
\newcommand{\NumVbMultiMax}{0.50}

\newcommand{\NumVbScoreMax}{0.61}
\newcommand{\NumVbScoreOver}{56}

\newcommand{\NumVbIrrA}{290}

\newcommand{\NumVbBoxA}{165}

\newcommand{\NumWbJointBig}{12}
\newcommand{\NumWbJointMax}{2.63}
\newcommand{\NumWbTermRegMax}{0.57}
\newcommand{\NumWbTermRegQ}{0.07}
\newcommand{\NumWbTermSmoothMax}{44}

\newcommand{\NumWbN}{68}

\newcommand{\NumWbCovWide}{106}

\newcommand{\NumRgCovA}{106/120}

\newcommand{\NumRgCovB}{70/80}

\newcommand{\NumRgCovC}{29/30}

\newcommand{\NumRgMaxDiff}{0.29}

\newcommand{\NumHetCavg}{$-1.92$}
\newcommand{\NumHetCrep}{$-0.27$}
\newcommand{\NumHetAngAvg}{131}
\newcommand{\NumHetAngRep}{105}
\newcommand{\NumHetCond}{$8.6\times10^{6}$}

\renewcommand{\NumVbCovHA}{90\%}

\renewcommand{\NumVbCoordHA}{89--96\%}
\renewcommand{\NumVbSdHA}{1.01--1.27}

\newcommand{\NumCalFlatCoverage}{94.4, 95.1 and 95.4}

\newcommand{\NumCalPetEach}{50}
\newcommand{\NumCalPetEstimatedCov}{98 and 98}
\newcommand{\NumCalPetOracleCov}{92 and 98}
\newcommand{\NumCalPetRawCov}{86 and 98}

\newcommand{\NumCalPetSeRatio}{0.94--1.08}
\newcommand{\NumCalPetSeRatioMed}{0.99 and 0.99}
\newcommand{\NumCalPetErrDiff}{0.28}

\newcommand{\NumCalPetDiscordant}{3 and 0}
\newcommand{\NumCalPetCovRelError}{41 and 40}

\makeatletter
\@ifundefined{acks}{\newenvironment{acks}[1][Acknowledgments]{\section*{#1}}{}}{}
\@ifundefined{funding}{\newenvironment{funding}{\section*{Funding}}{}}{}
\makeatother
\endlocaldefs
\AtBeginDocument{\hypersetup{pdftitle={Identification and curvature bias for shared inputs in multichannel dynamical models},pdfauthor={Hsin-Hsiung Huang},pdfsubject={Research preprint},pdfkeywords={curvature bias, incidental parameters, shared inputs, identifiability, dynamic PET}}}
\begin{document}
\begin{frontmatter}
\title{Identification and curvature bias for shared inputs in multichannel dynamical models}
\runtitle{Curvature bias for shared inputs}
\begin{aug}
\author[A]{\fnms{Hsin-Hsiung}~\snm{Huang}\ead[label=e1]{hsin.huang@ucf.edu}}
\address[A]{Dr. Yvette Kanouff School of Data, Mathematical, and Statistical Sciences, University of Central Florida, \printead{e1}}
\end{aug}
\begin{abstract}
Pooling many channels with a shared unknown input can leave persistent bias when each channel retains a fixed number of observations and its own incidental parameters. For heterogeneous smooth channel manifolds under isotropic noise, the second-order bias is an information-weighted regression of mean curvature on normal input velocity. A feasible correction gives centred Gaussian inference as the noise level $\sigma$ decreases with $R\sigma^6\to0$, compared with $R\sigma^2\to0$ without correction. We extend the stochastic expansion and studentization to deterministic channel arrays with densely correlated Gaussian errors of bounded covariance operator norm. General marginal covariance adds a forcing term that can bias even flat channels, while cross-channel covariance changes uncertainty. For Gaussian errors with a separable frame-by-channel covariance whose channel factor is autoregressive in a known ordering, an independent zero-mean calibration field yields a covariance estimator accurate enough for corrected inference. Boundary accumulation and fixed working population models introduce distinct bias terms; rational-multiplier and zero-count arguments connect identification to frame-integrated compartment models. Simulations test both oracle and estimated-covariance inference, including independent PET replications. Moderate-noise undercoverage, optimization ambiguity and the need to validate reconstruction covariance limit practical inference. The real-data example supports sensitivity analysis, not clinically calibrated intervals.
\end{abstract}
\begin{keyword}[class=MSC2020]
\kwd{Primary 62F12, 62F10; secondary 62J02, 62F30, 62P10}
\end{keyword}
\begin{keyword}
\kwd{blind deconvolution}\kwd{calibration}\kwd{compartment model}\kwd{identifiability}\kwd{incidental parameters}
\kwd{mean curvature}\kwd{Neyman--Scott problem}\kwd{positron emission tomography}
\end{keyword}
\end{frontmatter}

\section{Introduction}\label{sec:intro}
A collection of time series may be generated by different dynamical systems driven by the same unobserved input. In
dynamic positron emission tomography (PET), the tracer concentration in each region or voxel is the response of a
compartment system with its own rate constants to the arterial blood input. Total-body scanners now record such curves in
$10^5$--$10^6$ voxels \citep{vision_quadra}, and the input is usually taken from an image of the aorta; recent
likelihood-free and generative methods for voxelwise kinetic posteriors assume it is known
\citep{fan2021abc,gu2026vpetabc,zhao2025cm}. Voxelwise likelihoods that are nonlinear in a decay parameter also arise in positronium lifetime imaging with time-of-flight PET \citep{huang2025pli}. Estimating the input from the tissue curves themselves, by fitting all curves
jointly, is the idea behind simultaneous estimation (SIME) and multichannel blind identification
\citep{feng1997sime,wong2001sime,riabkov2002}. It is natural to expect that more channels give a better input. This paper
asks what pooling channels can recover. The answer has three parts: identification, the input scale, and estimation.

Identification requires more than pooling. An alternative input can move poles and zeros between its transform and the rational channel transfer functions. A binding input pole budget and suitable separation of channel poles and zeros rule out these alternatives up to scale. For frame-integrated exponential-polynomial models, including the delayed Feng input, a zero-count argument transfers continuous-curve identification and local rank to frame means under explicit endpoint, onset-chart and channel-type conditions (Section~\ref{sec:ident}). Structural identification does not guarantee good conditioning. An exact scale symmetry also remains: kinetic functionals that move with that symmetry retain external calibration error, however many channels are pooled.

The third part, and the principal statistical contribution, concerns estimation. Each channel brings its own kinetic
parameters, so the number of parameters grows in proportion to the number $R$ of channels while the number $n$ of time
frames stays fixed: the channel parameters are incidental in the sense of \citet{neyman1948}. We show that the joint
least-squares input estimator converges, as $R\to\infty$, to a pseudo-true input that differs from the truth by
\[
 -\frac{\sigma^2}2\bN^{-1}\bm+O(\sigma^4),\qquad
 \bN=\E\{B^T(I-P)B\},\quad \bm=\E\{s^2B^T(I-P)\kappa\},
\]
where $\sigma$ is the common noise level, $B$ is the sensitivity of a channel curve to the input, $I-P$ projects onto the normal
space of that channel's manifold of noise-free curves, and $(I-P)\kappa$ is the trace of the second fundamental form of
the manifold, $p$ times its mean curvature vector. The deviation is an information-weighted regression of curvature on the normal input
velocity. It is invariant under reparameterization of the channels, involves intrinsic but not parameter-effects curvature
\citep{bates1980}, and, under isotropic noise in the fitting metric, vanishes for flat channels such as unconstrained spectral basis models \citep{cunningham1993}. Because the
sampling standard deviation of the estimator is of order $\sigma/\sqrt R$, its bias dominates as soon as $R\sigma^2$ is
large: once $R\sigma^2\to\infty$, Wald intervals for every linear functional whose leading bias is nonzero cover with probability tending to zero. An analytic correction of the curvature term
removes the bias to first order and leaves the variance of the inverse fixed-effects information whenever $R\sigma^6\to0$. A PET simulation
with compact channel types and $\sigma$ decreasing along an explicit sequence in $R$ examines these predictions (Section~\ref{sec:numerics}). Physical
constraints in the joint fit change the order: channels on a boundary such as $k_3=0$ are projected onto a cone, and the bias becomes of
order $\sigma$ whenever their aggregate boundary forcing is nonzero.

A different approach is a population model, in which the channel parameters are random and a hierarchical posterior is computed for the input. With an
unrestricted population and support identification this gives consistency (Proposition~S8.1). With a working parametric population $g$ we show that the integrated
likelihood has its own second-order bias $-\sigma^2\bN^{-1}\E\{\operatorname{div}_\theta V+V^T\nabla\log g\}$, with
$-V=-GA^TB$ the velocity of the channel foot points; under Bernstein--von Mises conditions the same holds for the hierarchical posterior mean. The bias vanishes for the correct population, but it involves the population score in the directions of the weakly identified channels. In our
design the second-order coefficient is \NumPopRatiosText\ times the fixed-effects coefficient for Gaussian-mixture populations, although at realistic noise
levels the realized bias is damped well below this first-order value. The analytic correction, in contrast, needs no population model.

\paragraph{Scope and contributions}
The contribution lies in the combination of three features that template-estimation and incidental-parameter results do not cover jointly: the
channels are heterogeneous manifolds that move with the shared input, the number of frames per channel is fixed so that the bias is an expansion in
the noise level rather than in a per-unit sample size, and physical constraints make channels accumulate on boundary faces.
Sections~\ref{sec:ident}--\ref{sec:inference} connect frame-level identification to an invariant curvature-bias expansion and a feasible correction
at fixed frame count, and Section~\ref{sec:boundary} distinguishes boundary atoms from continuous accumulation near a face. Section~\ref{sec:population}
contrasts the population-score bias of integrated likelihood under a fixed working density. A covariance-aware extension (Section~\ref{sec:covariance}) permits dense Gaussian dependence, separates marginal bias from cross-channel uncertainty and gives an explicit sufficient covariance-calibration experiment. External calibration (Section~\ref{sec:minimax}) gives an additional limit. The numerical work tests the small-noise predictions in two prespecified
studies of a compact-population design, the second with the computed estimator (domain, input chart and regular-set correction) fixed in advance;
it separates bias from variance and from optimization error, records where the correction falls short at moderate noise, and illustrates the failure of two ad hoc averaging approximations in this heterogeneous design. The real-data example illustrates sensitivity rather than calibrated inference.

\paragraph{Related work}
Table~\ref{tab:novelty} summarizes the closest results. Inconsistency of maximum likelihood with incidental parameters goes back to \citet{neyman1948};
\citet{kiefer1956} showed that treating them as draws from an unknown distribution restores consistency, and \citet{lancaster2000} reviews the
econometric literature. Analytic and jackknife corrections for nonlinear panels \citep{hahn2004,arellano2007} and modified profile likelihoods
\citep{sartori2003} remove biases of order $1/T$ when each unit has $T\to\infty$ observations; here $n$ is fixed and the expansion is in the noise level.
Their large-$T$ guarantees do not directly cover this fixed-$n$ experiment. Splitting the frames can also remove the endpoint information required by our identification conditions; we do not establish that every possible jackknife construction must fail.
Small-noise biases of orthogonal-distance estimators are known in several special settings: the essential bias $\sigma^2/(2\varrho)$ of geometric
circle fits \citep{alsharadqah2009}, hyperaccurate corrections in geometric computer vision \citep{kanatani2013}, nonlinear orthogonal regression
\citep{fazekas2004} and principal curves \citep{hastie1989}. Closest to Theorem~\ref{thm:bias} is the template-estimation literature:
\citet{miolane2017} show that the Fr\'echet mean in a quotient by an isometric group action has asymptotic bias $-(\sigma^2/2)H$, with $H$ the trace of the second
fundamental form (unnormalized mean curvature vector) of the template orbit, and correct it by an iterative bootstrap; \citet{devilliers2017siims} and \citet{bigot2011} study inconsistency in related
settings. There, all observations lie on one orbit of a group that acts by isometries, so every observation carries the same geometry. Our channels are
different manifolds that move with the shared parameter, individual channels may be weakly informative, and the bias is the $\bN$-weighted regression of their
curvatures on their normal velocities. The scalar example below shows that the order of pooling and inversion matters even when every single-channel information matrix has condition number one.
The PET comparisons in Section~\ref{sec:numerics} are illustrative approximations, not implementations of the bootstrap methods in the template literature.
General-covariance geometric bias is also established \citep{kanatani1994,refregier1998}. Our extension derives a uniform correction and studentization theorem for heterogeneous fixed-frame arrays under Gaussian dependence; it does not claim the first covariance-dependent bias calculation. Boundary accumulation, population-score expansion and frame-level identification provide further results.
These distinctions specify the extension beyond that literature rather than a claim that curvature-induced bias itself is new. \citet{kumon1984} study the differential geometry of estimating functions with many nuisance parameters in
exponential families. Curvature in nonlinear regression and its effect on a single fit are classical \citep{box1971,bates1980,cook1986}; \citet{huang2023nonlinear} review the intrinsic and parameter-effects curvature measures and their effect on Wald inference. For regularized low-rank matrix models, \citet{huang2024lowrank} use a curvature bound for the fixed-rank manifold, proportional to the inverse smallest nonzero singular value, to control estimation error. Here the curvature enters the leading incidental-parameter bias. The algebraic
arguments relate to multichannel blind identification \citep{moulines1995,xu1995,abedmeraim1997}. For multi-region PET, \citet{holler2023} prove that
noiseless point measurements from several regions identify the irreversible two-tissue parameters and the input up to a common scale, with a consistency
result for Tikhonov-regularized estimation as the noise vanishes with a fixed number of regions, and \citet{holler2025reversible} extend this to
reversible kinetics. Section~\ref{sec:ident} overlaps with these results; it differs in using a rational-multiplier argument that covers a delayed
input with a repeated pole and a direct vascular term, in working with frame integrals, and in treating local rank. The statistical results, which
concern $R\to\infty$ channels with fixed frames, are not addressed there. Boundary parameters are well studied for a fixed number of parameters
\citep{self1987,andrews1999}; their accumulation in an incidental-parameter problem appears not to have been described. The empirical study of
\citet{ogden2010} found a systematic bias of SIME for [$^{18}$F]FDG and suggested that more regions with distinct kinetics might help;
Theorem~\ref{thm:bias} identifies a mechanism, present even under correct specification, that more channels at fixed noise cannot remove, and
Section~\ref{sec:numerics} shows that misspecification can produce larger shifts. The powers $R\sigma^2$ and $R\sigma^6$ follow by comparing bias orders
with the standard error; they are not asserted to be minimax phase boundaries.

\begin{table}[tb]
\caption{Closest results and what this paper adds.}\label{tab:novelty}
\centering\footnotesize
\begin{tabular}{@{}p{0.2\linewidth}p{0.36\linewidth}p{0.38\linewidth}@{}}
\toprule
Work & Setting and result & This paper \\
\midrule
Incidental-parameter panels \citep{hahn2004,arellano2007,sartori2003} & $T\to\infty$ observations per unit; bias $O(1/T)$, analytic and jackknife corrections & $n$ fixed; expansion in the noise level $\sigma$; regime $R\sigma^2$ \\
Geometric fitting \citep{alsharadqah2009,kanatani1994,kanatani2013} & points on one curve; essential bias $\sigma^2/(2\varrho)$ for circles & heterogeneous moving manifolds; covariance-aware correction with dense Gaussian dependence (Theorem~\ref{thm:covariance}) \\
Template estimation \citep{miolane2017} & isometric group action, all observations on one orbit; bias $-(\sigma^2/2)H$; bootstrap correction & non-isometric channel manifolds, individual channels may be weakly informative; bias $-\frac{\sigma^2}2\bN^{-1}\bm$ is an information-weighted regression; analytic one-step correction with $R\sigma^6\to0$; order-$\sigma$ bias with nonzero boundary forcing; population-score expansion \\
Multi-region PET identification \citep{holler2023,holler2025reversible} & noiseless point samples identify two-tissue kinetics and the input up to scale; Tikhonov consistency as noise vanishes with fixed regions & rational-multiplier proof with delay, repeated input pole and vascular term; frame integrals and local rank; inconsistency of joint estimation as $R\to\infty$ at fixed noise \\
SIME \citep{feng1997sime,wong2001sime,ogden2010} & joint estimation of input and kinetics; systematic bias observed empirically & a correctly specified bias mechanism that more channels cannot remove; misspecification shifts compared \\
Random effects \citep{kiefer1956} & unrestricted mixing distribution gives consistency & frame-level support condition (Proposition~S8.1); second-order bias of parametric populations (Theorem~\ref{thm:population}) \\
\bottomrule
\end{tabular}
\end{table}

\section{Multichannel experiments with a shared input}\label{sec:model}
For channel $r=1,\ldots,R$, consider the causal linear compartment system
\begin{equation}\label{eq:ode}
 \dot x_r(t)=\Omega(\theta_r)x_r(t)+b(\theta_r)u_\phi(t),\quad x_r(0)=0,
 \qquad f_r(t)=c(\theta_r)^Tx_r(t)+v_r u_\phi(t),
\end{equation}
with a matrix $\Omega(\theta_r)$ whose eigenvalues have nonpositive real parts, transfer function $H_r(z)=v_r+c(\theta_r)^T\{zI-\Omega(\theta_r)\}^{-1}b(\theta_r)$ and input $u_\phi$. For frame boundaries
$0=t_0<\cdots<t_n$ the observed frame means are
\begin{equation}\label{eq:data}
 y_r=F_r(\theta_r,\phi_0)+\sigma s_r\varepsilon_r\in\R^n,\qquad
 F_r(\theta,\phi)_i=\frac{w_{ri}}{t_i-t_{i-1}}\int_{t_{i-1}}^{t_i}f_r(t;\theta,\phi)\,dt,
\end{equation}
where $w_{ri}>0$ are known whitening weights, $s_r>0$ is a channel noise scale and $\varepsilon_r$ has identity covariance.
Channel types $\zeta_r=(\theta_r,s_r,w_r)$ are independent and identically distributed. The errors $\varepsilon_r$ are independent and identically distributed across channels and independent of all channel types. Section~\ref{sec:covariance} replaces this assumption by a specified Gaussian covariance model, allowing dependence between channels. We consider $R\to\infty$ with $n$ fixed and, in Section~\ref{sec:inference}, $\sigma\to0$ jointly with $R$.

\paragraph{Two-tissue channels with a Feng input}
Irreversible trapping permits a zero eigenvalue; asymptotic stability is therefore not required. Laplace transforms are taken in a right half-plane where they exist.
For a reversible two-tissue system with fractional blood volume $v\in(0,1)$,
\begin{equation}\label{eq:pet}
 H(z)=v+(1-v)K_1\frac{z+k_3+k_4}{z^2+(k_2+k_3+k_4)z+k_2k_4};
\end{equation}
here the same input drives tissue exchange and the vascular term. The irreversible model used for [$^{18}$F]fluorodeoxyglucose ([$^{18}$F]FDG) has $k_4=0$, a pole at $z=0$, and net influx rate $K_i=K_1k_3/(k_2+k_3)$.

\paragraph{Plasma and whole blood}
Tissue exchange is driven by the arterial plasma concentration $C_{\rm p}$, whereas the vascular term contains whole blood $C_{\rm b}$, and an aortic
image-derived input measures $C_{\rm b}$. If $C_{\rm b}=\varrho C_{\rm p}$ with constant $\varrho>0$, writing the model with plasma input gives vascular coefficient $v'=v\varrho$ and tissue coefficient $(1-v)K_1$. To keep the normalized form \eqref{eq:pet} one must also set
$K_1'=(1-v)K_1/(1-v\varrho)$, provided $v\varrho<1$; merely replacing $v$ is incorrect. Alternatively, using whole blood as the input retains $v$ and replaces $K_1$ by $K_1/\varrho$. The rates $k_2,k_3,k_4$ are unchanged, whereas $K_i$ depends on this convention and on calibration. If $v\varrho\ge1$, the plasma representation remains valid in free macro coefficients but not with a physical fractional blood volume. A time-varying ratio $\varrho(t)$ makes every channel a two-input system, outside the classes of
Section~\ref{sec:ident}; we treat it, together with reversible kinetics, channel-specific delays and dispersion, as a stress test in
Section~\ref{sec:numerics}, where the analysis model is deliberately wrong.
The Feng input \citep{feng1993input} is $u(t)=\{(A_1s-A_2-A_3)e^{\lambda_1s}+A_2e^{\lambda_2s}+A_3e^{\lambda_3s}\}\mathbf 1\{s>0\}$,
$s=t-\tau$, with $\lambda_j<0$ and transform
\[
U(z)=e^{-\tau z}\Bigl\{\frac{A_1}{(z-\lambda_1)^2}-\frac{A_2+A_3}{z-\lambda_1}+\frac{A_2}{z-\lambda_2}+\frac{A_3}{z-\lambda_3}\Bigr\}.
\]
For every $c>\max_rv_r$ the transformation
\begin{equation}\label{eq:scaleexact}
 \widetilde u=cu,\qquad \widetilde v_r=v_r/c,\qquad
 \widetilde K_{1r}=K_{1r}\frac{1-v_r}{c-v_r},
 \qquad \widetilde k_{jr}=k_{jr}\quad(j=2,3,4)
\end{equation}
leaves every output unchanged. Thus the input scale is not identified, the fraction $k_3/(k_2+k_3)$ and all rate constants
are scale free, and $K_i$ moves by the factor $(1-v_r)/(c-v_r)$. We remove the scale by a gauge, a slice $\Phi$ of input
parameters transversal to the orbits of \eqref{eq:scaleexact}; for the Feng input we fix $A_1$ and write
$\phi=(\tau,\log A_2,\log A_3,\log|\lambda_1|,\log|\lambda_2|,\log|\lambda_3|)$.
For statistical estimation we use positive amplitudes and an ordered chart, $\lambda_2<\lambda_3<0$, with all poles distinct; otherwise swapping $(A_2,\lambda_2)$ and $(A_3,\lambda_3)$ leaves the input unchanged. A sufficient physically positive subfamily has $\lambda_1<\lambda_2<\lambda_3<0$. Smooth local theory additionally restricts the delay to an interval strictly between two frame boundaries. Frame integration does not make the delayed onset six times differentiable across a boundary. These are restrictions of the statistical parameter space, not consequences of the algebraic theorem.

\section{Identification}\label{sec:ident}
\subsection{The admissible multiplier}
\begin{proposition}[Observational equivalence]\label{prop:multiplier}
Suppose $H_rU=\widetilde H_r\widetilde U$ for every channel, where all nonzero $H_r,\widetilde H_r$ are rational and
$U,\widetilde U$ are nonzero transforms. Then there is a nonzero rational function $Q$ with
\begin{equation}\label{eq:multiplier}
 \widetilde U=QU,\qquad \widetilde H_r=H_r/Q\quad (r=1,\ldots,R).
\end{equation}
Conversely, every $Q$ for which $QU$ and $H_r/Q$ remain in the specified input and channel classes produces identical
outputs. The transfer functions are identified up to scale exactly when every admissible multiplier is constant.
\end{proposition}

Supplement, Example~S1.1, gives four positive-system counterexamples illustrating the role of the input pole budget and common channel poles and zeros.

\subsection{Identification by pole separation}
For a set $\Lambda\subset\C$, write $\mathcal U_p(\Lambda)$ for the inputs whose transform is $e^{-\tau z}N(z)/D(z)$ with
$\tau\in\R$, $D$ monic of degree at most $p$ with all roots in $\Lambda$, and $N\ne0$ of degree less than $\deg D$; and
$\mathcal H_{n,q}$ for the rational channels $N/D$ with $D$ monic of degree at most $n$ and $\deg N\le q\le n$.
Representations are in lowest terms.
\begin{theorem}[Identification by pole separation]\label{thm:poles}
Let $U=e^{-\tau z}N_U/D_U\in\mathcal U_p(\Lambda)$ and nonzero $H_r=N_r/D_r\in\mathcal H_{n_r,q_r}$, $r=1,\ldots,R$, satisfy
\begin{enumerate}
\item[(i)] $\deg D_U=p$ and $\deg D_r=n_r$ for every $r$;
\item[(ii)] $\deg N_{r^*}=q_{r^*}$ for at least one channel $r^*$;
\item[(iii)] the polynomials $D_1,\ldots,D_R$ have no common root in $\Lambda$;
\item[(iv)] the polynomials $N_1,\ldots,N_R$ have no common root;
\item[(v)] $\gcd(N_U,D_r)=\gcd(N_r,D_U)=1$ for every $r$.
\end{enumerate}
If $\widetilde U\in\mathcal U_p(\Lambda)$ and $\widetilde H_r\in\mathcal H_{n_r,q_r}$ satisfy $\widetilde H_r\widetilde U=H_rU$ for every
$r$, then $\widetilde U$ has delay $\tau$, and there is a constant $c\ne0$ with $\widetilde U=cU$ and $\widetilde H_r=H_r/c$.
\end{theorem}
With $\Lambda=\C$ this is the usual pole-separation condition. Conditions (i)--(iv) cannot be dropped: Supplement Example~S1.1(b)
violates only the attainment of the input budget, (c) only (iii) and (d) only (iv); the supplement gives constructions for the
channel half of (i) and for (ii). With $R=1$, if the channel has a real pole $\rho\in\Lambda$ and $D_U$ a distinct real root $\rho_U$, the multiplier $Q=(z-\rho_U)/(z-\rho)$ is
admissible, so a single channel does not identify the input shape in these unrestricted rational classes. This construction need not preserve the narrower positive Feng family. The proof (Supplement~S1) compares the root multiplicities of
output denominators and numerators.

\begin{corollary}[Reversible two-tissue channels]\label{cor:feng}
Let $R\ge2$. Consider Feng inputs with $A_1A_2A_3\ne0$ and distinct $\lambda_1,\lambda_2,\lambda_3<0$ and channels \eqref{eq:pet}
with $(v_r,K_{1r},k_{2r},k_{3r},k_{4r})\in(0,1)\times(0,\infty)^4$. Outside a Lebesgue-null set of parameters, any Feng input and
channels of the form \eqref{eq:pet} that reproduce all outputs have the same delay and are related by \eqref{eq:scaleexact}.
\end{corollary}
\begin{corollary}[Irreversible two-tissue channels]\label{cor:irr}
The conclusion of Corollary~\ref{cor:feng} holds for channels \eqref{eq:pet} with $k_{4r}=0$ and $(v_r,K_{1r},k_{2r},k_{3r})\in
(0,1)\times(0,\infty)^3$, among alternative channels of the form \eqref{eq:pet} with $k_{4r}\ge0$; in particular, no Feng input with
reversible channels reproduces the outputs.
\end{corollary}
For irreversible channels all denominators $z(z+k_{2r}+k_{3r})$ share the root $0$, so condition (iii) fails with $\Lambda=\C$; it
holds with $\Lambda=\{\Re z<0\}$, which contains every Feng pole. The exceptional sets are zero sets of products of resultants,
nonzero at certified rational points (Supplement, Table~S1). The next example shows that restricting $\Lambda$ is
essential.
\begin{example}[Trapping versus a non-decaying input]\label{ex:trapping}
Let $u$ be any input and the channels irreversible with $v_r=0$. For $0<\gamma<\min_rk_{3r}$, the input
$\widetilde u(t)=u(t)+\gamma\int_0^tu$ and the reversible channels with
\[
 \widetilde K_{1r}=K_{1r},\quad \widetilde k_{2r}=k_{2r}+\gamma,\quad
 \widetilde k_{3r}=\frac{k_{2r}(k_{3r}-\gamma)}{k_{2r}+\gamma},\quad
 \widetilde k_{4r}=\frac{\gamma(k_{2r}+k_{3r})}{k_{2r}+\gamma}
\]
produce the same outputs; with $v_r>0$ an analogous map exists for small $\gamma$ (Supplement~S1.5). Here
$\widetilde U(z)=U(z)(z+\gamma)/z$ has a pole at $0$.
\end{example}
Within the Feng class the exact equivalence is excluded. A slowly decaying Feng component can nevertheless resemble a persistent component over a finite scan, suggesting weak numerical separation between slow input dynamics and channel reversibility; no uniform approximation of the full transformed input by the Feng family is claimed.
For the study analysed in Supplement~S30, $1/|\lambda_3|=\NumLamThreeInv$\,min exceeds the scan length.

\subsection{Finite frames}\label{sec:frames}
Theorem~\ref{thm:poles} and its corollaries compare complete curves, whereas the data are frame means. For two-tissue channels and the Feng
input both inputs and outputs are exponential polynomials, and a classical zero count transfers identification from curves to frames. A real
exponential polynomial is $E(t)=\sum_jp_j(t)e^{\mu_jt}$ with distinct real $\mu_j$ and real polynomials $p_j\not\equiv0$. Its order
$\sum_j(\deg p_j+1)$ is the order of the smallest constant-coefficient differential equation it solves. A nonzero exponential polynomial
of order $M$ has at most $M-1$ real zeros, counted with multiplicity \citep[Part V, Problem 75]{polya1925}. Say that a model has
\emph{exponential orders} $(p_U,p_H)$ if every input vanishes before its onset $\tau\ge0$ and is $u(\tau+s)=g(s)$ for $s\ge0$, with $g$ of order at most $p_U$, real exponents and $g(0)=0$, and
every channel has impulse response $v\delta_0+\kappa$ with $\kappa$ of order at most $p_H$ and real exponents. Then the channel curve is
$f_r(\tau+s)=h_r(s)=v_rg(s)+(\kappa_r*g)(s)$, which we continue to all real $s$ as an exponential polynomial. The Feng input has $p_U=4$. Two-tissue
channels, reversible or irreversible, have real poles and $p_H=2$, and so do the macro channels of Section~\ref{sec:numerics}.
\begin{theorem}[Finite frames]\label{thm:frames}
Consider a model with exponential orders $(p_U,p_H)$, delays in $[\tau_-,\tau_+]$, and frame endpoints that include at least
$m^*=2(p_U+p_H)+1$ points in $[\tau_+,\infty)$.
\begin{enumerate}
\item[(a)] If two parameter values give equal frame means in channel $r$, their curves agree on $[\tau_+,\infty)$ and $\widetilde h_r(s)=h_r(s+\widetilde\tau-\tau)$ for
all real $s$.
\item[(b)] Suppose further that one true curve is positive on $(0,\tau_+-\tau_-]$ and that the continuations $h_1,\ldots,h_R$ have no common zero in
$[-(\tau_+-\tau_-),0)$. Then equal frame means in all channels imply $\widetilde\tau=\tau$ and equal complete curves. In particular, the conclusions of
Theorem~\ref{thm:poles} and Corollaries~\ref{cor:feng}--\ref{cor:irr} hold for frame means, against alternatives in the same model of exponential orders.
\item[(c)] On an open real-analytic parameter chart with locally bounded real exponents and onset strictly inside a fixed frame, for any tangent direction $\xi$, the derivative of the frame means along $\xi$ vanishes if and
only if the derivative of every complete curve vanishes at all times other than its onset.
\end{enumerate}
\end{theorem}
The proof (Supplement~S4) applies the zero count to differences of cumulative curves, which frame means determine. Part (b)
excludes an earlier alternative onset whose curves continue the true curves backwards. When alternative curves are nonnegative, it suffices,
in place of the common-zero condition, that one true channel has $v_r>0$ and $g'(0)>0$, because that curve then rises with positive slope at its onset.
The bound $m^*$ is not claimed to be sharp. Part (c) reduces local rank for frames to an
infinitesimal version of Theorem~\ref{thm:poles}. In the next proposition $\delta U$ and $\delta H_r$ denote first-order variations of the
transforms, and the delay variation enters through $\delta U=e^{-\tau z}\{-z\,\delta\tau\,N_U/D_U+\delta(N_U/D_U)\}$.
\begin{proposition}[Infinitesimal identification]\label{prop:infinitesimal}
Assume conditions (i)--(v) of Theorem~\ref{thm:poles}, and consider variations tangent to the classes: $\deg\delta D_U<\deg D_U$,
$\deg\delta N_U\le\deg N_U$, $\deg\delta D_r<n_r$ and $\deg\delta N_r\le q_r$. If $H_r\delta U+U\delta H_r=0$ for every $r$, then $\delta\tau=0$,
$\delta D_U=0$, $\delta D_r=0$, and $\delta N_U=cN_U$, $\delta N_r=-cN_r$ for a constant $c$.
\end{proposition}
Two degree conditions carry the delay step. The input numerator must keep its degree, and the channel $r^*$ of condition (ii) must attain its numerator
degree. For the Feng input, $\deg N_U=2$ exactly when $g'(0)=A_1+A_2(\lambda_2-\lambda_1)+A_3(\lambda_3-\lambda_1)\ne0$, which holds for positive amplitudes when
$\lambda_1<\lambda_2,\lambda_3$. For two-tissue and macro channels, the numerator degree is attained exactly when $v_r\ne0$. If either condition fails, a
delay change can be absorbed by the amplitudes.
\begin{corollary}[Frame-level conditions for two-tissue channels]\label{cor:framespet}
Let the input be Feng with $A_1$ fixed, positive amplitudes, $\lambda_1<\lambda_2<\lambda_3<0$, and delays in $[\tau_-,\tau_+]$. Let the channels be
reversible or irreversible two-tissue channels with $v_r\in(0,1)$ and positive rates, or macro channels with $v_r,a_{1r},a_{2r}\ne0$ and $\beta_r>0$.
Suppose at least 13 frame endpoints lie in $[\tau_+,\infty)$, and the condition of Theorem~\ref{thm:frames}(b) holds. For every set of channels containing two
channels outside the exceptional null sets of Corollaries~\ref{cor:feng}--\ref{cor:irr}, defined for macro channels by the same products of
resultants in macro coordinates:
\begin{enumerate}
\item[(i)] frame means determine the delay, the input shape and every channel among Feng inputs in the chart and channels with at most two real poles (in the rational classes of Theorem~\ref{thm:poles}), in particular the unconstrained macro extension;
\item[(ii)] the stacked frame Jacobian in $(\phi,\theta_1,\ldots,\theta_R)$ has full column rank. Hence every channel Jacobian $\partial_\theta F_r$ has full
column rank, and the information matrix $\sum_rB_r^T(I-P_r)B_r$ of Section~\ref{sec:bias} is positive definite.
\end{enumerate}
Consequently, for a channel-type law on a compact set of such types that charges every neighbourhood of two generic types, and whose support contains a
type with continued curve negative on $[-(\tau_+-\tau_-),0)$ and a type with curve positive on $(0,\tau_+-\tau_-]$, parts (c) and (d) of
Assumption~\ref{ass:reg} in Section~\ref{sec:bias} hold: each channel is identified given the input, $\bN\succ0$, and $Q_0$ vanishes only at $\phi_0$.
\end{corollary}
The 35 frames of the study of Supplement~S30 place 13 endpoints after every onset up to \NumFrLatest\,min and \NumFrEndAfter\ at or after the end of the
frame containing the reference onset, so the delay chart is that frame, of width $L=1/12$\,min. For condition (b) an analytic certificate replaces
pointwise evaluation. If the input continuation satisfies $g'>0$ on $[-L,0]$, then $g<0$ and $|g|$ is decreasing towards the onset there, and for an
irreversible two-tissue channel with $b=k_2+k_3$ the continuation obeys
\begin{equation}\label{eq:certificate}
 h(s)\le-\bigl\{v-(1-v)K_1L\,e^{bL}\bigr\}\,|g(s)|\qquad(-L\le s<0),
\end{equation}
where $\kappa(x)=(1-v)K_1(k_3+k_2e^{-bx})/b$ is the continued tissue impulse response, $0<\kappa(x)\le(1-v)K_1e^{bL}$ on $[-L,0]$, and the convolution
is continued by the oriented integral, so that the tissue part $(\kappa*g)(s)=\int_0^s\kappa(s-t)g(t)\,dt=-\int_s^0\kappa(s-t)g(t)\,dt$ is nonnegative and at most
$(1-v)K_1e^{bL}L|g(s)|$. Hence $h<0$ on $[-L,0)$
whenever $v>(1-v)K_1Le^{bL}$, and one such channel type excludes a common zero. For positive Feng amplitudes with $\lambda_1<\lambda_2<\lambda_3<0$, $g'>0$ for $s\le0$ and $g>0$ for $s>0$ follow directly from the exponential terms (Supplement~S4.4). The certificate holds
for four of the eight organ parameter sets of the XCAT phantom used in Section~\ref{sec:numerics}, for
instance the lungs, where $(1-v)K_1Le^{bL}=0.020<v=0.087$ (Supplement~S4.4). It concerns the simulation population, not the voxels.
Under these conditions the identification and rank parts of Assumption~\ref{ass:reg} follow for the actual frames; the smoothness part still requires the
delay to stay inside one frame.

Corollary~\ref{cor:framespet} is a statement of structural identification: no two parameter values give the same frame means. It does not bound
the conditioning, which governs estimation. On the compact set of channel types of the corollary, $\lambda_{\min}(A_r^TA_r)$ is a continuous positive function of the type, and $\lambda_{\min}(\bN)$
a continuous positive function of the delay over a closed subinterval of the chart, so both are bounded below uniformly; but the bounds can be small. For the eight XCAT organ means and delays across the
interior of the chart, $\lambda_{\min}(\bN)$ ranges over \NumCondLminRange\ and the condition number of $\bN$ over \NumCondRange, while
$\min_r\lambda_{\min}(A_r^TA_r)\ge\NumCondAtA$ (Supplement~S4.4). Weak numerical identification, a nearly flat criterion in a
direction that is identified in principle, is therefore expected along the slow input components, and it is distinct from structural
nonidentification, for which the criterion would be exactly flat.

\subsection{Information for the input shape and scale}
Consider the exact whitened tangent experiment $Y_r=A_ra_r+B_rh+\varepsilon_r$, $Z=Ch+\xi$, with independent standard
Gaussian noises, unrestricted channel nuisance $a_r$, and common input perturbation $h\in\R^q$. Let $P_r$ be the orthogonal
projection onto $\operatorname{col}(A_r)$ and $N_R=\sum_{r=1}^RB_r^T(I-P_r)B_r$.
\begin{proposition}[Profile information]\label{prop:schur}
The efficient information for $h$ is $I_h=C^TC+N_R$. If $B_rh_0\in\operatorname{col}(A_r)$ for every $r$ and $Ch_0=0$, then
$h_0$ is unidentifiable for every $R$.
\end{proposition}
The scale symmetry \eqref{eq:scaleexact} differentiates into a vector $e$ with $B_re\in\operatorname{col}(A_r)$ for all $r$.
\begin{proposition}[Shape information grows, scale information does not]\label{prop:shape}
Suppose $B_re\in\operatorname{col}(A_r)$ for every $r$, for a unit vector $e$, and $I_h$ is invertible. For the efficient least-squares estimator of Proposition~\ref{prop:schur}, $N_Re=0$ and
\begin{equation}\label{eq:scalefloor}
 \operatorname{var}(g^T\widehat h)=g^TI_h^{-1}g\ \ge\ \frac{(g^Te)^2}{\norm{Ce}^2}\qquad(g\in\R^q,\ R\ge1).
\end{equation}
If $(A_r,B_r)$ are independent copies of $(A,B)$ with $\E\norm B_{\mathrm F}^2<\infty$ and $\bN=\E\{B^T(I-P_A)B\}$ is positive
definite on $e^\perp$, then $R\,g^TI_h^{-1}g\to g^T\bN^+g$ for $g\perp e$, with $\bN^+$ the Moore--Penrose inverse, and $e^TI_h^{-1}e\to\norm{Ce}^{-2}$.
\end{proposition}
Its restriction to the gauge slice $\Phi$ reappears below as the curvature of the profiled criterion.

\section{Incidental channels and the curvature bias}\label{sec:bias}
\subsection{The joint estimator and its pseudo-true value}
The joint, or fixed-effects, least-squares estimator solves
\begin{equation}\label{eq:joint}
 (\widehat\phi_R,\widehat\theta_1,\ldots,\widehat\theta_R)\in\argmin_{\phi\in\Phi,\ \theta_r\in\Theta}\ \sum_{r=1}^R\norm{y_r-F_r(\theta_r,\phi)}^2 .
\end{equation}
It is the estimator of SIME and of joint blind identification. Profiling the channels,
$\widehat\phi_R\in\argmin_\phi\sum_r\rho_r(y_r,\phi)$ with
\begin{equation}\label{eq:rho}
 \rho_r(y,\phi)=\min_{\theta\in\Theta}\norm{y-F_r(\theta,\phi)}^2=\operatorname{dist}^2\{y,\mathcal M_r(\phi)\},\qquad
 \mathcal M_r(\phi)=F_r(\Theta,\phi).
\end{equation}
The channel manifold $\mathcal M_r(\phi)$ is the set of noise-free curves that channel $r$ can produce with input $\phi$. Write
$\rho(y,\phi)$, $F(\theta,\phi)$ for a generic channel and $Q_\sigma(\phi)=\E\rho(y,\phi)$ for the expected criterion.

At the true curve $x_r=F_r(\theta_r,\phi_0)$ define $A_r=\partial_\theta F_r$ ($n\times p$), $B_r=\partial_\phi F_r$ ($n\times q$),
$G_r=(A_r^TA_r)^{-1}$, $P_r=A_rG_rA_r^T$ and
\begin{equation}\label{eq:curv}
 \kappa_r=\sum_{j,k=1}^p(G_r)_{jk}\,\partial^2_{\theta_j\theta_k}F_r,\quad
 h_r=(I-P_r)\kappa_r,\quad N_r=B_r^T(I-P_r)B_r,\quad m_r=B_r^Th_r .
\end{equation}
In the coordinates $\theta$, the second fundamental form of $\mathcal M_r(\phi_0)$ at $x_r$ is
$\mathrm{II}(a,b)=(I-P_r)\partial^2_\theta F_r[a,b]$, and $h_r=\sum_{jk}(G_r)_{jk}\mathrm{II}(e_j,e_k)$ is its trace with respect to the
induced metric $A_r^TA_r$: $p$ times the mean curvature vector. It depends on the manifold only. The tangential part $P_r\kappa_r$
measures parameter-effects curvature \citep{bates1980}. The columns of $(I-P_r)B_r$ are the normal velocities of the channel
manifold when the input moves. Put $\bN=\E N_r$ and $\bm=\E(s_r^2m_r)$.

\begin{assumption}\label{ass:reg}
\begin{enumerate}
\item[(a)] $1\le p<n$, $q\ge1$, and $\Theta\subset\R^p$ and $\Phi\subset\R^q$ are compact; the channel types have compact support, the true channel parameters
lie in a compact subset $\Theta_0$ of the interior of $\Theta$, and $\phi_0$ is interior to $\Phi$.
\item[(b)] $F_r(\theta,\phi)=F(\theta,\phi;w_r)$, where $F$ is six times continuously differentiable in $(\theta,\phi)$ on a
neighbourhood of $\Theta\times\Phi$, with derivatives bounded uniformly in $w$ and continuous in $w$.
\item[(c)] (Channels given the input.) $\lambda_{\min}(A_r^TA_r)\ge c_0>0$, and for every $\eta>0$,
$\inf\{\norm{F_r(\theta,\phi_0)-x_r}:\theta\in\Theta,\ \norm{\theta-\theta_r}\ge\eta\}>0$, uniformly over channel types.
\item[(d)] (Input.) $Q_0(\phi)=\E\operatorname{dist}^2\{x,\mathcal M(\phi)\}$ vanishes on $\Phi$ only at $\phi_0$, and $\bN$ is
positive definite.
\item[(e)] (Noise.) $\varepsilon_r$ is symmetric about zero, $\E\varepsilon_r\varepsilon_r^T=I_n$ and $\E\exp(t_0\norm{\varepsilon_r}^2)<\infty$ for some $t_0>0$.
\end{enumerate}
\end{assumption}
Condition (d) is global identification in the finite-frame noise-free experiment: $Q_0(\phi)=0$ means that almost every true channel curve can be
reproduced with input $\phi$. Condition (c) asks that each channel be identified once the input is known. The continuous-curve theorem alone does not
give either condition for frames. For Feng inputs in the ordered chart and two-tissue or macro channels, Corollary~\ref{cor:framespet} derives both
from the algebra under its stated conditions on the frames, the delay chart and the type law; the endpoint count alone does not suffice. For other
models, (c) and (d) remain assumptions. The macro
parameterization extends the physical channel manifold across $v=0$ and $k_3=0$, so true channels with $v,k_3>0$ are interior. Section~\ref{sec:boundary}
shows that imposing the physical boundaries in the joint fit changes the order of the bias. The data-generating law of the validation design in Section~\ref{sec:numerics} admits these conditions on suitable compact parameter sets; the computed fits are assessed separately from the exact constrained minimizers; the reference design, with unbounded log-normal
channel draws, does not satisfy them literally, and the experiments report how often fits reach the solver box.

\begin{theorem}[Curvature bias]\label{thm:bias}
Under Assumption~\ref{ass:reg}, for every $\sigma>0$ the function $Q_\sigma$ is continuous on $\Phi$ and, almost surely, every limit
point of $\widehat\phi_R$ lies in $\argmin Q_\sigma$. There are $\sigma_1>0$ and $C<\infty$ such that for $0<\sigma\le\sigma_1$:
\begin{enumerate}
\item[(i)] every $\bar\phi\in\argmin Q_\sigma$ satisfies
\begin{equation}\label{eq:bias}
 \Bigl\|\bar\phi-\phi_0+\frac{\sigma^2}2\bN^{-1}\bm\Bigr\|\le C\sigma^4;
\end{equation}
\item[(ii)] uniformly over a neighbourhood of $\phi_0$,
\begin{equation}\label{eq:qexp}
 Q_\sigma(\phi)=Q_0(\phi)+\sigma^2\E\bigl[s^2\bigl\{n-p-\ip{x-\pi_\phi(x)}{h_\phi(\pi_\phi x)}\bigr\}\bigr]
 +O(\sigma^4+\sigma^2\norm{\phi-\phi_0}^2),
\end{equation}
where $\pi_\phi(x)$ is the point of $\mathcal M(\phi)$ nearest to the noise-free curve $x$ and $h_\phi$ is the curvature vector
\eqref{eq:curv} of $\mathcal M(\phi)$.
\end{enumerate}
\end{theorem}
The criterion \eqref{eq:qexp} explains the mechanism. Without noise, every input $\phi$ is penalized by the squared distance
$Q_0(\phi)$ of the true curves from its channel manifolds, which is quadratic with Hessian $2\bN$ at $\phi_0$. Noise adds
$\sigma^2(n-p)\E s^2$, the part of the noise normal to the manifold, minus a correction that is linear in the displacement
$x-\pi_\phi(x)$ and in the curvature: a manifold that bends towards the data captures more of the noise. The minimizer therefore
moves in the direction in which the true curves come to lie on the concave side of their manifolds, by an amount set by the
regression of the curvature on the normal input velocities. Symmetry of the noise removes the odd orders; with skewed noise the
remainder in \eqref{eq:bias} is $O(\sigma^3)$, with a leading term given by the third moments (Supplement, Remark~S3.4).

\begin{proposition}[Geometry of the bias]\label{prop:geometry}
Under Assumption~\ref{ass:reg}:
\begin{enumerate}
\item[(a)] (Invariance.) If $\theta=g(\vartheta)$ for a $C^6$ diffeomorphism $g$, the profiled criterion $\sum_r\rho_r$, hence
$\widehat\phi_R$ and $\argmin Q_\sigma$, are unchanged, and so are $h_r$, $N_r$ and $m_r$.
\item[(b)] (Flat channels.) Let $\Phi$ be compact, $F_r(\theta,\phi)=a_r(\phi)+X_r(\phi)\theta$ with $\Theta=\R^p$ and $X_r(\phi)$ of full
column rank uniformly on $\Phi$, with $a_r,X_r$ continuous and bounded, and let $\varepsilon_r$ have mean zero and identity covariance; assume also $\E\norm{x_r}^2<\infty$ and $\E s_r^2<\infty$; parts (a), (c) and (e) of Assumption~\ref{ass:reg} are not needed. Then $Q_\sigma(\phi)=Q_0(\phi)+\sigma^2(n-p)\E s^2$ for every $\sigma$; if $Q_0$
vanishes only at $\phi_0$, $\widehat\phi_R\to\phi_0$ almost surely.
\item[(c)] (Minimal channels.) If $h_r=0$ almost surely, the $\sigma^2$ term of \eqref{eq:bias} vanishes.
\item[(d)] (Channel bias.) For $0<\sigma\le\sigma_1$ and $\bar\phi\in\argmin Q_\sigma$, the channel estimate with the limiting input,
$\theta^*(y_r,\bar\phi)$, the minimizer near $\theta_r$ of $\norm{y_r-F_r(\theta,\bar\phi)}$, satisfies
\begin{equation}\label{eq:chanbias}
 \E\{\theta^*(y_r,\bar\phi)\mid\zeta_r\}-\theta_r=-\frac{\sigma^2}2G_rA_r^T\bigl\{s_r^2P_r\kappa_r-B_r\bN^{-1}\bm\bigr\}+O(\sigma^4),
\end{equation}
where the expectation is over the event, of probability $1-O(e^{-c/\sigma^2})$, on which the minimizer is defined.
\end{enumerate}
\end{proposition}
The first term of \eqref{eq:chanbias} is Box's coordinate-dependent bias of a single nonlinear regression \citep{box1971,cook1986}. Suitable local coordinates can eliminate it at a chosen point; a single global reparameterization need not remove it everywhere, and nonlinear target functionals can still be biased. The second is the input bias propagated through the
channel, which depends on the intrinsic curvature of all channels and cannot. For the two-tissue model in its macro form
$F=v\,C+a_1(C*1)+a_2(C*e^{-\beta t})$, the channel is linear in $(v,a_1,a_2)$ and only the direction of $\beta=k_2+k_3$ curves, so
the curvature that matters is that of the curves as the washout rate changes; spectral models with fixed rates and unconstrained
coefficients are flat and immune, whereas the nonnegativity constraints of \citet{cunningham1993} and \citet{gunn2002} create
boundaries, which change the order of the bias (Theorem~\ref{thm:boundary}).

\begin{example}[Known special cases]\label{ex:known}
(a) \emph{Circle fitting.} Points $y_r=(a+\varrho\cos\theta_r,b+\varrho\sin\theta_r)+\sigma\varepsilon_r$ with angles uniform on the circle
have $n=2$, $p=1$, $\phi=(a,b,\varrho)$, $N_r=v_rv_r^T$ and $m_r=-v_r/\varrho$ with $v_r=(\cos\theta_r,\sin\theta_r,1)^T$, so
$\bN^{-1}\bm=-(0,0,1)^T/\varrho$ and the radius of the geometric fit converges to $\varrho+\sigma^2/(2\varrho)+O(\sigma^4)$, the essential bias
of \citet{alsharadqah2009}. (b) \emph{Template orbits.} If $y_r=g_r\phi+\sigma\varepsilon_r$ with $g_r$ in a compact Lie group acting
orthogonally on $\R^n$, $\phi_0$ has trivial stabilizer, and the template is parameterized in a neighbourhood of $\phi_0$ in the normal
slice $\phi_0+E\psi$ of its orbit, then $N_r=I_q$ and $m_r=E^Th_0$, with $h_0$ the curvature vector of the orbit at $\phi_0$, and
\eqref{eq:bias} becomes $-(\sigma^2/2)E^Th_0$, Theorem~2 of \citet{miolane2017}. Both are
verified in Supplement, Section~S3.4.
\end{example}
The two cases use symmetry differently. For template orbits every channel sees the same geometry, $N_r=I_q$ and $m_r=E^Th_0$, so the bias is that
of a single orbit. In circle fitting the channel quantities are not identical: $N_r=v_rv_r^T$ has rank one and $m_r=-v_r/\varrho$ depends on the angle,
so no single point identifies the circle. All points lie on one manifold, however, and the rotations about the centre map the channel quantities into
one another, so with uniform angles $\bN$ and $\bm$ are averages over a single orbit of the group. With heterogeneous channels the leading bias $-\frac{\sigma^2}2(\sum_rN_r)^{-1}\sum_rs_r^2m_r$ is generally not the unweighted average of single-type biases. Individual $N_r$ may be singular or poorly conditioned. Global ambiguity in the unrestricted rational class does not imply singular local information in the narrower Feng family: a single channel can have full local rank while retaining a discrete global ambiguity. In the design of Section~\ref{sec:numerics}, at the eight XCAT organ means, the single-organ matrices $N_r$ have condition numbers between $4\times10^3$ and $10^8$, so single-type biases are unstable or undefined, whereas the eight types together have condition number $1.2\times10^4$. The proof of Theorem~\ref{thm:bias} therefore cannot use the group structure or the quotient on
which these arguments rest. It works instead with one profiled criterion that is uniform over channel types.

\paragraph{A well-conditioned reversal under pooling}
Let $n=2$, $p=q=1$, $s_r=1$ and
\[
 F(\theta,\phi)=\bigl(\theta,\ g(\theta)+(1+\theta)\phi\bigr)^T,
 \qquad g(\theta)=\theta^4/4-5\theta^3/3+2\theta^2.
\]
Take $\phi_0=0$ and $\theta_r=0$ or $1$ with equal probabilities, inside the compact channel set $[-1/4,5/4]$ and a compact input interval containing zero in its interior.
At the two true types, $A=(1,0)^T$, $(N,m)=(1,4)$ and $(4,-6)$, respectively. Thus the pooled coefficient is
$-\tfrac12\bar N^{-1}\bar m=1/5$, whereas the average of individual coefficients is
$-\tfrac12\E(N^{-1}m)=-5/8$. Every scalar $N$ has condition number one. The first output identifies $\theta$ and $1+\theta>0$ identifies $\phi$, so the example satisfies the identification conditions.
The reversal is caused by information weighting, without singular inverses. It illustrates a distinction between two algebraic operations, not the performance of an existing competing estimator.

\subsection{Inference and the curvature correction}\label{sec:inference}
The sampling fluctuation of $\widehat\phi_R$ is of order $\sigma/\sqrt R$ and its bias of order $\sigma^2$, so the ratio is
$\sqrt R\sigma$. Let $\bN_s=\E(s_r^2N_r)$ and $\Sigma=\bN^{-1}\bN_s\bN^{-1}$; for $s_r\equiv1$, $\Sigma=\bN^{-1}$.
\begin{theorem}[Local asymptotics]\label{thm:local}
Let Assumption~\ref{ass:reg} hold, $R\to\infty$ and $\sigma=\sigma_R\to0$ with $\sigma_R^2\log R\to0$. Then, for every
$\bar\phi_R\in\argmin Q_{\sigma_R}$,
\begin{equation}\label{eq:clt}
 \frac{\sqrt R}{\sigma_R}(\widehat\phi_R-\bar\phi_R)\rightsquigarrow N_q(0,\Sigma).
\end{equation}
Consequently, if $\sqrt R\sigma_R\to\lambda\in[0,\infty)$, then $(\sqrt R/\sigma_R)(\widehat\phi_R-\phi_0)\rightsquigarrow N_q(-\tfrac\lambda2\bN^{-1}\bm,\Sigma)$; if
$\sqrt R\sigma_R\to\infty$, then $\sigma_R^{-2}(\widehat\phi_R-\phi_0)\to-\frac12\bN^{-1}\bm$ in probability. For $c\in\R^q$ with $c^T\bN^{-1}\bm\ne0$,
the interval $c^T\widehat\phi_R\pm z_{1-\alpha/2}\widehat v_c\sigma_R/\sqrt R$ with $\widehat v_c^2\to c^T\Sigma c$ in probability has
asymptotic coverage $\Prob\{|Z+\lambda c^T\bN^{-1}\bm/(2\sqrt{c^T\Sigma c})|\le z_{1-\alpha/2}\}$, $Z\sim N(0,1)$, which is $1-\alpha$ only
for $\lambda=0$ and tends to zero as $\lambda\to\infty$.
\end{theorem}
For fixed $\sigma\le\sigma_1$ with $c^T\bN^{-1}\bm\ne0$ ($\sigma_1$ reduced if needed), every limit point of $c^T\widehat\phi_R$ differs
from $c^T\phi_0$ by at least $\frac{\sigma^2}4|c^T\bN^{-1}\bm|$ (Theorem~\ref{thm:bias}), so every interval centred at $c^T\widehat\phi_R$ with width $o_P(1)$ has coverage
tending to zero. Over all linear functionals,
the largest standardized bias is $\sqrt R\sigma\,\gamma$ with $\gamma=\frac12(\bm^T\bN_s^{-1}\bm)^{1/2}$, which is $\frac12(\bm^T\bN^{-1}\bm)^{1/2}$ when $s_r\equiv1$
(Supplement~S3.5). In the design of Section~\ref{sec:numerics}, $\gamma=\NumGamma$, so the worst functional is biased by two
standard errors once $R\ge\NumRcritOne$ at $\sigma=0.05$ and $R\ge\NumRcritTwo$ at $\sigma=0.2$.

The bias in \eqref{eq:bias} is computable from the fit. Let $\widehat\theta_r=\theta^*(y_r,\widehat\phi_R)$,
$\widehat\sigma_r^2=\rho_r(y_r,\widehat\phi_R)/(n-p)$, and let $\widehat N$ and $\widehat m$ be the averages of $N_r$ and
$\widehat\sigma_r^2m_r$ evaluated at $(\widehat\theta_r,\widehat\phi_R)$; when $s_r\equiv1$, $\widehat\sigma_r^2$ may be replaced by the pooled
estimate. The curvature-corrected estimator is
\begin{equation}\label{eq:corrected}
 \widetilde\phi_R=\widehat\phi_R+\tfrac12\widehat N^{-1}\widehat m .
\end{equation}
It needs the first and second channel derivatives and the first input derivatives of $F$ at the fit, using dense linear algebra at cost $O\{R(np^2+npq+nq^2+p^3) + q^3\}$ once derivatives are available; forward evaluation and differentiation costs are additional.
\begin{theorem}[Curvature correction]\label{thm:corrected}
Under the conditions of Theorem~\ref{thm:local}, if in addition $R\sigma_R^6\to0$, then
\[
 \frac{\sqrt R}{\sigma_R}(\widetilde\phi_R-\phi_0)\rightsquigarrow N_q(0,\Sigma).
\]
For Gaussian noise with $s_r\equiv1$, $\sigma^2(\sum_rN_r)^{-1}$ is the inverse of the efficient information for $\phi$ when
$\theta_1,\ldots,\theta_R$ are treated as unknown constants (Proposition~\ref{prop:schur}), and the asymptotic variance of $\widetilde\phi_R$
equals it to first order; no optimality is claimed among procedures that exploit the distribution of the channel types.
\end{theorem}
The correction extends valid inference from $R=o(\sigma^{-2})$ channels to $R=o(\sigma^{-6})$; the proof is in
Supplement~S3.6. At moderate noise higher-order terms matter: Supplement~S12 gives a finite-design expansion of
the next terms, and Supplement~S7 a conditional antithetic bootstrap correction for them, whose conditions are not verified for the
PET model.

\paragraph{The computed estimator}
Theorems~\ref{thm:local} and~\ref{thm:corrected} concern the exact minimizer over the compact sets $\Theta$ and $\Phi$ of
Assumption~\ref{ass:reg}(a), and a computation must fix them. In Section~\ref{sec:numerics}, $\Theta$ is a solver box and $\Phi$ an ordered input
chart, both chosen before the second validation study. Near degenerate faces of a channel family ($\beta\to0$ and $\beta\to\infty$ for the macro
channel) $A_r^TA_r$ is nearly singular and $m_r$, which contains $G_r$, is unbounded, so a few fitted channels can dominate $\widehat m$ while carrying
little information about $\phi$. We therefore use the regular-set correction
\begin{equation}\label{eq:regcorr}
 \widetilde\phi^{\rm reg}_R=\widehat\phi_R+\tfrac12\widehat N^{-1}\widehat m_{\rm reg},\qquad
 \widehat m_{\rm reg}=\frac1R\sum_{r\in\widehat{\mathcal G}_R}\widehat\sigma_r^2\,m_r(\widehat\theta_r,\widehat\phi_R),
\end{equation}
where $\widehat{\mathcal G}_R$ contains the channels with $\widehat\theta_r$ interior to $\Theta$ and $\lambda_{\min}(\widehat A_r^T\widehat A_r)\ge\kappa$ for a
fixed $\kappa>0$. If $\kappa<c_0$, every channel is regular on events of probability tending to one, so
$\Prob(\widetilde\phi^{\rm reg}_R=\widetilde\phi_R)\to1$ and Theorem~\ref{thm:corrected} holds for $\widetilde\phi^{\rm reg}_R$
(Supplement, Corollary~S3.5). Unlike $\widetilde\phi_R$, the regular set depends on the channel coordinates. This is an asymptotic equivalence, not a finite-sample improvement: the two estimators differ only through
fitted channels outside the simultaneous regular-projection event, and a large difference warns against relying on the local approximation. The assumptions concern the true channels and do not rule out such fitted channels at finite noise. The domain is part of the estimator,
so enlarging $\Theta$ defines a different estimator, which Section~\ref{sec:numerics} examines as such.

On a smooth branch, interval estimates use the sandwich variance of the profiled estimating equation, $\widehat H^{-1}(\sum_r\widehat\psi_r\widehat\psi_r^T)\widehat H^{-1}$ with
$\widehat H$ the profile Hessian of the criterion, with independent channels as sampling units (Supplement~S3.5). For half the criterion the channel score is $-B_r^Te_r$, with $e_r$ the whitened residual. Replacing it by $-B_r^T(I-P_r)e_r$ requires the nuisance normal equations; at an active box face the full-tangent projection generally changes the score (Supplement~S23). Grouping frames
within a voxel does not account for spatial dependence between voxels. Replacing $\widehat H$ by its Gauss--Newton form $2\sum_rN_r(\widehat\theta_r,\widehat\phi_R)$
is asymptotically equivalent in the correctly specified small-noise regime, but the two differ at moderate noise (Section~\ref{sec:numerics}); neither is a
fixed-noise misspecification-robust sandwich.

\subsection{Covariance changes both bias and uncertainty}\label{sec:covariance}
The curvature-only formula requires isotropic errors in the fitting metric. General covariance also couples tangential noise to the motion of the normal space. This matters when diagonal frame weights do not fully whiten a reconstruction. With isotropic marginal noise, dependence within independent clusters of fixed size changes only the score covariance, not the curvature bias (Supplement, Theorem~S10.1); but independence of arbitrarily chosen image blocks is not guaranteed. We therefore give a deterministic-design extension allowing dense Gaussian dependence; it is not a model-free justification for reconstructed PET intervals.

Write $Q_r=I-P_r$, $H_{r,jk}=\partial_{\theta_j\theta_k}^2F_r$ and $V_r=G_rA_r^TB_r$. For a positive definite marginal covariance $C_r$ define
\begin{align}
 M_r^C&=G_rA_r^TC_rA_rG_r,\quad
 \kappa_r^C=\sum_{jk}(M_r^C)_{jk}H_{r,jk},\notag\\
 D_{r,j}&=\partial_{\theta_j}B_r-\sum_kH_{r,jk}(V_r)_{k,:},\notag\\
 t_r(C_r)&=\tfrac12B_r^TQ_r\kappa_r^C-
       \sum_jD_{r,j}^T(Q_rC_rA_rG_r)_{:,j}.\label{eq:covforce}
\end{align}
Here $D_{r,j}$ is an $n\times q$ matrix. When $C_r=s_r^2I$, the second term vanishes and $t_r=s_r^2m_r/2$. Although its two terms use coordinates, their sum is invariant under smooth channel reparameterization (Supplement~S13).

\begin{theorem}[Gaussian dependence and covariance-aware correction]\label{thm:covariance}
Let the channel types form deterministic triangular arrays satisfying the uniform compactness, smoothness and channel separation in Assumption~\ref{ass:reg}(a)--(c). Replace (d) by uniform separation of $R^{-1}\sum_r\rho(x_r,\phi)$ away from $\phi_0$, and assume $N_R=R^{-1}\sum_rN_r\to N\succ0$. Let
$y_r=x_r+\sigma\xi_r$, where $\xi=(\xi_1^T,\ldots,\xi_R^T)^T\sim N(0,K_R)$,
$C_r=(K_R)_{rr}\succeq cI$ and $\|K_R\|_{\rm op}\le\bar c$, for constants $c,\bar c>0$.
Put $L_R=(B_1^TQ_1,\ldots,B_R^TQ_R)$, $t_R=R^{-1}\sum_rt_r(C_r)$ and assume
$\Omega_R=R^{-1}L_RK_RL_R^T\to\Omega\succ0$.
If $\sigma\to0$ and $\sigma^2\log R\to0$, the exact joint least-squares estimator satisfies
\begin{equation}\label{eq:covexp}
 \widehat\phi_R-\phi_0=\frac{\sigma}{R}N_R^{-1}L_R\xi
 -\sigma^2N_R^{-1}t_R+O_P(\sigma^4+\sigma^2/\sqrt R).
\end{equation}
Evaluate hatted geometry at the joint fit and set
\[
 \widehat\sigma^2=\frac{\sum_r\rho(y_r,\widehat\phi_R)}{\sum_r\tr(\widehat Q_rC_r)},\quad
 \widehat t_R=R^{-1}\sum_rt_r(C_r;\widehat\theta_r,\widehat\phi_R),\quad
 \widetilde\phi_C=\widehat\phi_R+\widehat\sigma^2\widehat N_R^{-1}\widehat t_R.
\]
If also $R\sigma^6\to0$, then
$\sqrt R(\widetilde\phi_C-\phi_0)/\sigma\Rightarrow N_q(0,N^{-1}\Omega N^{-1})$.
The consistent covariance estimate is
$\widehat\sigma^2\widehat N_R^{-1}\widehat\Omega_R\widehat N_R^{-1}/R$,
where $\widehat\Omega_R=R^{-1}\widehat L_RK_R\widehat L_R^T$.
The same conclusions hold for a regular-set correction with any fixed $0<\kappa<c_0$.
\end{theorem}
With $K_R=\operatorname{diag}(s_r^2I_n)$ and known scales bounded away from $0$ and $\infty$, the forcing reduces to $s_r^2m_r/2$ and the expansion and limits agree with those of Theorems~\ref{thm:local} and~\ref{thm:corrected} for deterministic channel arrays with Gaussian noise; the corrected estimators coincide when $s_r\equiv1$ and the pooled noise scale is used. The proof combines a uniform profile expansion with a Gaussian Poincar\'e bound; the leading score is exactly Gaussian, so no artificial partition into independent spatial clusters is needed. For $K_R=S_R\otimes C$ with $(S_R)_{rs}=\varrho^{|r-s|}$ and fixed $|\varrho|<1$, its operator norm is bounded. Conversely, unrestricted long-range dependence need not satisfy the theorem. Consistent estimation of $K_R$ is an additional problem: operator-norm consistency suffices for the variance, but replacing the $C_r$ in the bias correction also requires
$\sqrt R\sigma\max_r\|\widehat C_r-C_r\|_{\rm op}\to_P0$ (Supplement, Corollary~S13.2).

\begin{example}[Bias without curvature]\label{ex:flatcov}
Let $F(\theta,\phi)=(\theta,\theta\phi)^T$, $\phi_0=0$, $C=\left(\begin{smallmatrix}1&a\\a&1\end{smallmatrix}\right)$ with $0<|a|<1$, and $R^{-1}\sum_r\theta_r^2\to m_2>0$. The channels are straight lines, but $t_r(C)=-a$ and the leading bias is $a\sigma^2/m_2$. For unrestricted line projections the population minimizer is exactly
$2a\sigma^2/\{m_2+(m_2^2+4a^2\sigma^4)^{1/2}\}$.
Spatial correlation changes the variance through $K_R$ while leaving this marginal expected criterion unchanged. Thus a variance adjustment alone does not correct the bias. Correct marginal generalized least squares also removes this population bias; this example does not establish superiority over a correctly specified fitting metric.
\end{example}

\begin{proposition}[A sufficient covariance-calibration experiment]\label{prop:covcalibration}
Under Theorem~\ref{thm:covariance}, suppose $K_R=S_R(\varrho)\otimes C$, where $S_R(\varrho)_{rs}=\varrho^{|r-s|}$, $|\varrho|<\varrho_*<1$, $C\succ0$ is fixed and $\tr C=n$. Observe $M_R\ge1$ independent calibration fields $Z^{(m)}\sim N(0,\tau_R^2K_R)$, independent of the fitted data, with arbitrary unknown $\tau_R>0$. Define
\[
 W_R=\frac{\sum_{m,r}Z^{(m)}_r Z^{(m)T}_r}{M_RR},\quad
 \widehat C=\frac{nW_R}{\tr W_R},\quad
 \widehat\varrho=\operatorname{clip}_{[-\varrho_*,\varrho_*]}
 \frac{\sum_m\sum_{r<R}Z^{(m)T}_rZ^{(m)}_{r+1}}{\sum_m\sum_{r<R}\|Z^{(m)}_r\|^2}.
\]
Then $\widehat K_R=S_R(\widehat\varrho)\otimes\widehat C$ satisfies
$\|\widehat K_R-K_R\|_{\rm op}=O_P\{(M_RR)^{-1/2}\}$.
Replacing the covariance in Theorem~\ref{thm:covariance} retains its corrected limit and studentization, even for $M_R=1$.
\end{proposition}
The proof in Supplement~S13.5 uses Gaussian quadratic-form variances and an operator-norm Lipschitz bound uniform in $R$. Covariance estimation changes the standardized correction by $O_P(\sigma/\sqrt{M_R})$. One calibration field contains $R$ correlated channel vectors; it is additional information, not the residuals of a single fitted image. The covariance structure and known zero mean are substantive assumptions.

\subsection{Physical constraints and boundary channels}\label{sec:boundary}
Kinetic parameters obey sign constraints such as $k_3\ge0$ and $v\ge0$, and basis-function methods impose nonnegativity
\citep{cunningham1993,gunn2002}. If these constraints are imposed in the joint fit, a channel whose true parameter lies on or near a boundary is
projected onto a cone rather than a manifold. Consider one linear constraint $\theta_1\ge0$ per channel, with the other coordinates interior. Let
$a_r=\partial_{\theta_1}F_r$, let $\bar a_r=(I-P_{r,-})a_r$ be its part orthogonal to the other columns of $A_r$, and put
\[
 e_r=\bar a_r/\norm{\bar a_r},\qquad c_r=B_r^Te_r\in\R^q,\qquad d_r=\norm{\bar a_r}\theta_{r1}\ge0 .
\]
Thus $c_r$ is the input velocity along the constrained direction, and $d_r$ is the linearized distance of the true curve from the face of its channel manifold,
in the metric of the whitened data; $\norm{\bar a_r}^2=1/\{(A_r^TA_r)^{-1}\}_{11}$ is bounded below under Assumption~\ref{ass:reg}(c). Write
$x_-=\min(x,0)$, $x_+=\max(x,0)$, $\varphi_{\rm N}$ and $\Phi_{\rm N}$ for the standard normal density and distribution function, and
$\psi(v)=\varphi_{\rm N}(v)-v\{1-\Phi_{\rm N}(v)\}=\E(Z-v)_+$ for $Z\sim N(0,1)$.
\begin{theorem}[Boundary channels]\label{thm:boundary}
Let Assumption~\ref{ass:reg} hold for the extended model with Gaussian noise, and let the true channels lie in $\{\theta_1\ge0\}$, possibly on the face $\{\theta_1=0\}$. Let $\widehat\phi^+_R$ be the joint estimator with $\theta_{r1}\ge0$ in every channel, and let $Q^+_\sigma$ be its expected
criterion. Then every limit point of $\widehat\phi^+_R$ lies in $\argmin Q^+_\sigma$. For $0<\sigma\le\sigma_1$, every $\bar\phi^+\in\argmin Q^+_\sigma$ satisfies
\[
 \norm{\bar\phi^+-\phi_0-\sigma\delta_\sigma}\le C\sigma^2,
\]
where $\delta_\sigma$ is the unique root of $\Gamma_\sigma(\delta)=\bN\delta+\E[s_rc_r\psi\{(d_r/\sigma-c_r^T\delta)/s_r\}]$. Moreover
$\norm{\delta_\sigma}\ge\norm{\Gamma_\sigma(0)}/\lambda_{\max}(\bN+\E c_rc_r^T)$, and $\Gamma_\sigma(0)\to(2\pi)^{-1/2}\E\{s_rc_r\mathbf 1(d_r=0)\}$ as $\sigma\to0$.
\end{theorem}
If the aggregate boundary forcing $\E\{s_rc_r\mathbf 1(d_r=0)\}$ is nonzero, the constrained joint estimator is therefore
biased by an amount of exact order $\sigma$, against $\sigma^2$ without the constraint. A positive fraction of channels on the face is necessary for this but
not sufficient, because boundary velocities of different channel types can cancel.
\begin{example}[Cancelling boundary forcing]\label{ex:cancel}
Let $n=2$, $p=q=1$, $F_c(\theta,\phi)=(\theta+c\phi,\ \phi)^T$ with $\theta\ge0$, and let the channel types $c\in\{-1,1\}$ be equally likely, all with
$\theta_r=0$, $s_r=1$ and $\phi_0=0$. Every channel lies on the face, $\bar a_r=e_r=(1,0)^T$, $c_r=c$, $d_r=0$ and $\bN=1$. Then
$\Gamma_\sigma(\delta)=\delta+\frac12\{\psi(-\delta)-\psi(\delta)\}=\frac32\delta$, because $\psi(-v)-\psi(v)=v$, so $\delta_\sigma=0$. Indeed the reflection $(\phi,c,y_2)\mapsto(-\phi,-c,-y_2)$ leaves the model and the noise law invariant, so the constrained estimator has no bias of order
$\sigma$ although every channel is at the boundary.
With $c_r$ uniform on $\{1,2\}$ instead, $\E c_r\ne0$ and $\delta_\sigma=-0.301$. Because the model is linear, $\bar\phi^+=\sigma\delta_\sigma$ exactly here. In 20 simulated data sets with $R=2\times10^5$ channels at each of
$\sigma=0.02$, 0.05 and 0.1 (\texttt{check\_boundary\_example.py}), the constrained estimate divided by $\sigma$ averaged $-0.301$ in the second case and was
within $5\times10^{-4}$ of zero in the first.
\end{example}
When the forcing is nonzero, the order-$\sigma$ bias is large relative to the nominal fixed-effects standard error, which is of order
$\sigma/\sqrt R$: the standardized bias grows like $\sqrt R$ at every noise level, and the curvature correction does not remove it. Channels
farther than $\sigma\{s_r(2\log(1/\sigma))^{1/2}+|c_r^T\delta_\sigma|\}$ from the face contribute only to the $\sigma^2$ term. Newton's method from $\delta=0$ gives
$\delta_\sigma\approx-\widetilde N_\sigma^{-1}\Gamma_\sigma(0)$ with $\widetilde N_\sigma=\bN+\E\{\Phi_{\rm N}(-d_r/(\sigma s_r))c_rc_r^T\}$; this one-step value is accurate only
when $\norm{\Gamma_\sigma(0)}$ is small relative to $\lambda_{\min}(\bN)$, and we solve for the root in the experiments (proof in
Supplement~S5). Several active constraints per channel lead to a projection onto a polyhedral cone and a
multivariate truncated-normal forcing term; we do not treat that case.

The forcing term has a direct reading. For the unconstrained fit the linearized slack estimate is $X_r=d_r+\sigma s_r\zeta_r$, and
$\E(X_r)_-=-\sigma s_r\psi\{d_r/(\sigma s_r)\}$. Thus $-\sigma\Gamma_\sigma(0)$ is the average of the expected constraint violations of the unconstrained fits,
weighted by the boundary velocities. More generally $\E\{(c_r^T\delta-X_r/\sigma)/s_r\}_+=\psi\{(d_r/\sigma-c_r^T\delta)/s_r\}$, so the unconstrained fits estimate
$\Gamma_\sigma$ without bias to first order under correct specification. Violations by unconstrained voxel fits, as in Supplement~S30, therefore
indicate the forcing term that a constrained input fit would create; on misspecified data the indication is only qualitative. These results suggest a two-stage rule. First estimate the shared input with the smooth
extension, to which Theorems~\ref{thm:bias}--\ref{thm:corrected} apply. Then impose the physical constraints only when each channel is fitted
with the estimated input. In the second stage the boundary is a fixed-dimensional problem for each channel \citep{self1987,andrews1999}, and its
error does not accumulate over channels. A parametric bootstrap of the constrained input fit is not expected to be a remedy, because the local parameters $d_r/\sigma$ cannot be estimated
consistently \citep{andrews2000}; we have not studied it.

Continuous accumulation near a boundary differs from an atom on the face. If its slack density behaves as $d^{\alpha-1}$, $0<\alpha<1$, a nonzero leading boundary coefficient gives bias of order $\sigma^{1+\alpha}$. Supplement, Proposition~S11.1, states the precise regularity and noncancellation conditions and derives the coefficient. This dominates smooth curvature bias. A bounded slack density instead gives an $O(\sigma^2)$ displacement, whose coefficient can still differ from the unconstrained one.

\section{Population models for the channels}\label{sec:population}
An unrestricted mixing distribution can restore consistency under compactness and support identification (Supplement, Proposition~S8.1), following \citet{kiefer1956}. This does not assert consistency for a fixed parametric population. Here the channels are independent, $s_r\equiv1$ and the noise is Gaussian. For a fixed working density $g$ on $\Theta$, let
\[
 \ell_r(\phi;g)=\log\int_\Theta\varphi_\sigma\{y_r-F_r(\theta,\phi)\}g(\theta)\,d\theta,
\]
and let $\widehat\phi_g$ maximize $\sum_r\ell_r(\phi;g)$. Write $V_r=G_rA_r^TB_r$ and let $\operatorname{div}_\theta V_r$ denote columnwise divergence.
\begin{theorem}[Population-score bias]\label{thm:population}
Let Assumption~\ref{ass:reg} hold with Gaussian noise and $s_r\equiv1$, let $\Theta$ be the closure of an open set with Lipschitz boundary (a box, for
instance), and strengthen part (b) to $F\in C^8$ on a neighbourhood, with uniformly bounded derivatives. Let $g$ be positive and $C^6$ on a neighbourhood of $\Theta$, with bounded derivatives and $\inf_\Theta g>0$.
Every limit point of $\widehat\phi_g$ maximizes $L_\sigma(\phi)=\E\ell_r(\phi;g)$, and there are $\sigma_1>0$ and $C<\infty$ such that for $0<\sigma\le\sigma_1$ every
maximizer $\bar\phi_g$ satisfies
\begin{equation}\label{eq:popbias}
 \bigl\|\bar\phi_g-\phi_0+\sigma^2\bN^{-1}t_g\bigr\|\le C\sigma^4,\qquad
 t_g=\E\bigl\{\operatorname{div}_\theta V_r+V_r^T\nabla\log g(\theta_r)\bigr\}.
\end{equation}
\end{theorem}
The proof uses a uniform Laplace expansion and a volume identity that combines the profile curvature with the Laplace determinant (Supplement~S9). For a smooth true population $g_0$ supported in the interior, integration by parts gives $\E\{\operatorname{div}V+V^T\nabla\log g_0\}=0$: the coefficient measures population-score misfit in the foot-point directions $V$. The exact correctly specified likelihood also has the truth as an expected maximizer at every noise level, by the Kullback--Leibler inequality. This last fact does not follow by substituting $g_0$ into the theorem, whose positive-density assumptions differ.

Supplement~S8 gives precise results for correct, flat and estimated populations. A jointly fitted working family needs additional uniformity and an identified limiting population; a hierarchical posterior conclusion further requires Bernstein--von Mises and moment conditions, unverified here. In the XCAT design, fixed Gaussian-mixture working populations leave second-order coefficients \NumPopRatiosText\ times the fixed-effects coefficient in the information norm. These coefficients concern a small-noise expansion, not realized finite-noise bias. The analytic correction needs no population model. The attempted hierarchical sampler did not converge and remains a supplementary diagnostic, not evidence for posterior calibration.

\section{Calibration limits}\label{sec:minimax}
An exact scale symmetry leaves absolute kinetic targets dependent on external calibration, however many channels are pooled (Proposition~\ref{prop:shape}). Supplement, Proposition~S15.1, gives a van Trees bound along that symmetry orbit. With Gaussian log-scale calibration error of variance $\tau^2$, its lower bound for aggregate squared log-influx error is $R\tau^2/(1+\pi^2\tau^2/\delta^2)$ on a regular orbit interval of half-width $\delta$. This scale limitation is separate from the shape bias and from covariance-shape calibration in Proposition~\ref{prop:covcalibration}.

\section{Numerical experiments}\label{sec:numerics}
\paragraph{Common setting}
All PET simulations use the 35 frames of the total-body study of Supplement~S30 (12$\times$5\,s, 4$\times$15\,s, 6$\times$1\,min,
5$\times$2\,min, 4$\times$3\,min, 4$\times$5\,min) and, as true input, the Feng fit to its aortic image-derived input
($\tau=0.40$\,min, $\lambda=(-26.8,-0.227,-0.0123)$\,min$^{-1}$; relative error of the fit \NumIdifFit). Channel parameters
$(K_1,k_2,k_3,v)$ are built from the eight [$^{18}$F]FDG organ parameter sets of the XCAT phantom distributed with
\texttt{qurit/dynamicPET} \citep{qurit_dynamicpet}, for myocardium, liver, lungs, kidneys, spleen, bone, white and grey matter, with $k_4=0$. An organ is chosen
uniformly, its blood-volume value is floored at 0.005, and each log parameter is perturbed independently. The noise of frame $i$ in
channel $r$ has standard deviation $\sigma\bar f_r u_i$, where $\bar f_r$ is the mean of the channel curve and $u_i^2\propto\bar c_i/\Delta_i$
follows the population-mean curve $\bar c$ and the frame duration, normalized so that $\sigma$ is a coefficient of variation; the
data are whitened by these known shapes, so $s_r\equiv1$. Channels are fitted in the macro parameterization $(v,a_1,a_2,\log\beta)$; near the true curves it has the
same channel manifold, so the same $h_r$, $N_r$ and $m_r$ (Proposition~\ref{prop:geometry}(a)), and it extends the manifold across the
boundaries $v=0$ and $k_3=0$. For every candidate input each channel is fitted by linear least squares in $(v,a_1,a_2)$ on a grid of $\beta$ followed by
Levenberg--Marquardt refinement; the input is updated by Gauss--Newton steps on the profiled criterion, and the channel search is repeated at the
final input until no channel improves. The procedure does not certify a global optimum, so every fit reports its standardized profile score (about zero at a stationary point, one
at a point one standard error away); Supplement~S16.2 states conditions under which the limit theory transfers to computed estimates,
and Supplement~S31 records seeds, run times and failures.

\paragraph{Computational provenance}
An earlier implementation of the frame-integrated convolution was inaccurate when a channel washout rate nearly coincided with an input decay rate.
The repaired kernel agrees with 50-digit quadrature to \NumRkMaxErr\ in \NumRkChecks\ checks, and every result in this section was computed with it;
earlier-kernel computations are historical analyses (Supplement~S26), not used for any conclusion.

\paragraph{Designs}
The \emph{validation design} has a compact data-generating law compatible with Theorems~\ref{thm:local} and~\ref{thm:corrected}. Each log
parameter is perturbed by $0.2z$ with $z$ standard normal truncated to $[-2,2]$, so the channel types have compact support, interior to the solver box
$\log\beta\in[-6,3]$ ($\beta\in[0.0025,20]$\,min$^{-1}$); the certificate \eqref{eq:certificate} holds for every lung-type channel, so
Corollary~\ref{cor:framespet} gives Assumption~\ref{ass:reg}(c)--(d). No estimator is started at the truth. Each data set has its own simulated
image-derived input, the true input dispersed by an exponential kernel of mean 5\,s with 5\% multiplicative frame noise; its Feng fit starts the joint
estimator, a median of \NumValStartMed\ and at most \NumValStartMax\ sandwich standard errors from the truth in the worst coordinate. The channel grid has
361 points, and each fit ends with a search along the two weakest directions of the information for a lower local minimum
(Supplement~S19). The \emph{reference design} (unbounded log-normal perturbations, a 157-point grid, true start) is used below only for
physical constraints; the \emph{stress design} generates data from deliberately different models.

\paragraph{The predicted bias}
In the coordinates $(\tau,\log A_2,\log A_3,\log|\lambda_1|,\log|\lambda_2|,\log|\lambda_3|)$ the predicted deviation per unit $\sigma^2$ in the validation
design is $-\frac12\bN^{-1}\bm=$ \NumValBpred, with $\gamma=\NumValGamma$; $\bN$ has condition number $\NumValCondN$, and
$\lambda_{\min}(A_r^TA_r)\ge\NumValAtA$ in $5000$ draws. The two weakest directions of $\bN$ are made of the amplitudes and decay rates of the two slow
input components, which the trapping term can partly mimic (Example~\ref{ex:trapping}). We summarize an estimate by its standardized projection
$c=b^T\bN(\widehat\phi-\phi_0)/(b^T\bN b)$ on the predicted direction $b=-\frac12\bN^{-1}\bm$, fixed in advance; the population minimizer has
$c=\sigma^2+O(\sigma^4)$. This functional attains the largest standardized bias $\sqrt R\sigma\gamma$, so Theorem~\ref{thm:local} predicts that nominal
95\% intervals for it from the joint estimator cover with probability about $\Phi(1.96-\sqrt R\sigma\gamma)-\Phi(-1.96-\sqrt R\sigma\gamma)$, whereas
Theorem~\ref{thm:corrected} predicts 95\% for the corrected estimator. Theorem~\ref{thm:corrected} concerns distributional centring, not a finite-sample
expectation, so empirical means assess higher-order terms, finite $R$ and computation together.

\paragraph{First validation study}
The noise level decreases along $\sigma_R=0.2(R/256)^{-1/4}$ for $R=256$, 1024 and 4096, so that $R\sigma_R^2=10$, 20 and 41 grows while
$R\sigma_R^6\le0.016$. The study comprises all \NumValNA, \NumValNB\ and \NumValNC\ planned data sets, fixed before the runs and generated from disjoint,
prespecified random-number streams; no computation failed (Table~\ref{tab:validation}). The joint fit had $c/\sigma_R^2=\NumValProjJA$, \NumValProjJB\
and \NumValProjJC\ (Monte Carlo standard errors \NumValProjJAse, \NumValProjJBse, \NumValProjJCse) and its intervals covered in \NumValCovJA, \NumValCovJB\
and \NumValCovJC\ of the data sets, against \NumValPredA, \NumValPredB\ and \NumValPredC\ predicted; the bias exceeded its first-order value by 13--16\%.
The smooth correction \eqref{eq:corrected} had $c/\sigma_R^2=\NumValProjCA$, \NumValProjCB\ and \NumValProjCC, a residual bias of 0.5--0.7 standard
errors, and its intervals covered in \NumValCovCA, \NumValCovCB\ and \NumValCovCC\ (exact intervals \NumValCovCACI, \NumValCovCBCI, \NumValCovCCCI). The
correction thus removed most of the bias but did not deliver nominal coverage for the projection at $R=256$ and 1024. Coordinatewise the shortfall was
larger: at $R=256$ the corrected intervals for the six coordinates covered in \NumDecCovLoA--\NumDecCovHiA\ of the data sets, and the empirical spread
exceeded the Gauss--Newton sandwich standard error by factors \NumDecSdLoA--\NumDecSdHiA; at $R=1024$ and 4096 the coverage was
\NumDecCovLoB--\NumDecCovHiB\ and \NumDecCovHiC\ (Supplement, Table~S11 and Figure~S6).

\begin{table}[tb]
\caption{Validation design, regenerated with the repaired convolution kernel (all planned data sets, independent seeds fixed in advance):
compact channel law, $\sigma_R=0.2(R/256)^{-1/4}$ and a simulated image-derived start. Mean standardized projection $c/\sigma^2$ on the
predicted bias direction (Monte Carlo standard error), mean error divided by the mean sandwich standard error, coverage of nominal 95\% intervals
with exact 95\% intervals, the first-order coverage predicted by Theorem~\ref{thm:local} (joint) or Theorem~\ref{thm:corrected} (corrected), and the
ratio of the empirical standard deviation to the mean sandwich standard error with a paired bootstrap interval.}
\label{tab:validation}
\centering\footnotesize\setlength{\tabcolsep}{3.5pt}
\resizebox{\linewidth}{!}{\begin{tabular}{rrrlrrrrrr}
\toprule
$R$ & $\sigma_R$ & $R\sigma_R^2$ & estimator & $c/\sigma^2$ & bias/se & covered & 95\% CI (\%) & predicted (\%) & sd/se\\
\midrule
256 & 0.200 & 10.2 & joint & 1.14 (0.06) & 2.31 & 43/120 & [27, 45] & 46 & 1.26 [1.11, 1.39]\\
256 & 0.200 & 10.2 & corrected & 0.26 (0.05) & 0.54 & 106/120 & [81, 93] & 95 & 1.21 [1.07, 1.35]\\
\addlinespace
1024 & 0.141 & 20.5 & joint & 1.16 (0.04) & 3.41 & 6/80 & [3, 16] & 17 & 1.03 [0.89, 1.15]\\
1024 & 0.141 & 20.5 & corrected & 0.22 (0.04) & 0.64 & 70/80 & [78, 94] & 95 & 1.02 [0.88, 1.13]\\
\addlinespace
4096 & 0.100 & 41.0 & joint & 1.13 (0.04) & 4.68 & 0/30 & [0, 12] & 1 & 0.88 [0.68, 1.04]\\
4096 & 0.100 & 41.0 & corrected & 0.16 (0.04) & 0.65 & 29/30 & [83, 100] & 95 & 0.88 [0.68, 1.04]\\
\bottomrule
\end{tabular}}
\end{table}

\paragraph{Bias and variance at smaller noise}
A prespecified experiment with $R=\NumSnR$ and \NumSnPairs\ channel samples, each analysed at $\sigma=0.2$, 0.1 and 0.05 with antithetic noise
$\pm\varepsilon$ and the same channels, noise array and start, separated residual bias from variance (Table~\ref{tab:smallnoise}). Half-sums of
antithetic pairs contain the bias but no first-order noise: the joint estimator had $c/\sigma^2=\NumSnJoint$ and the corrected one \NumSnCorr\ (Monte
Carlo standard errors \NumSnCorrSe), a corrected bias of \NumSnCorrBias\ standard errors. At fixed $R$ the second-order bias of the fitted estimator is not
the population coefficient, and a finite antithetic average retains random quadratic terms; Supplement~S12 derives these finite-design
coefficients. Their realized values average 1.039 (Monte Carlo s.e.\ 0.021) for the joint fit and 0.035 (0.020) for the correction; the remainders after
subtracting them fall by factors of 3--5 per halving of $\sigma$, which explains most of the apparent discrepancy from 1 and 0 at small noise. Half-differences retain the first-order noise; their spread was \NumSnSdse\ times the sandwich standard error along the
projection and \NumSnSdseCoordBy\ times coordinatewise. At $R=\NumSnR$, then, the residual is higher-order bias that shrinks with $\sigma$, and the
sandwich is accurate for the first-order noise term.

\begin{table}[tb]
\caption{Bias and variance at decreasing noise ($R=1024$, compact channel law, antithetic noise pairs with common channel
samples, noise and simulated image-derived starts across $\sigma$). Half-sums cancel odd noise terms and estimate the mean bias: mean standardized projection $c/\sigma^2$
(Monte Carlo standard error over pairs) of the joint and corrected estimators and the corrected bias in units of the mean sandwich standard error.
Half-differences retain odd terms, led by the first-order noise: their root mean square over the mean sandwich standard error, along the projection and over the six
coordinates. Coverage (\%) of nominal 95\% sandwich intervals over both members of every pair, along the projection (joint, corrected) and the
range over the six coordinates (corrected).}
\label{tab:smallnoise}\centering\footnotesize\setlength{\tabcolsep}{3pt}
\resizebox{\linewidth}{!}{\begin{tabular}{rrccccccc c}\toprule
 & & \multicolumn{3}{c}{bias (half-sums)} & \multicolumn{2}{c}{noise / se (half-differences)} & \multicolumn{3}{c}{coverage (\%)}\\
\cmidrule(lr){3-5}\cmidrule(lr){6-7}\cmidrule(lr){8-10}
$\sigma$ & pairs & joint $c/\sigma^2$ & corrected $c/\sigma^2$ & corr.\ bias/se & projection & coordinates & joint & corrected & corr.\ coord.\\\midrule
0.2 & 30 & 1.19 (0.02) & 0.32 (0.02) & 1.30 & 0.95 & 0.70--1.27 & 0 & 73 & 68--98\\
0.1 & 30 & 1.09 (0.02) & 0.11 (0.02) & 0.23 & 1.01 & 0.77--1.10 & 48 & 97 & 88--100\\
0.05 & 30 & 1.05 (0.02) & 0.05 (0.02) & 0.05 & 0.99 & 0.76--1.09 & 80 & 98 & 87--100\\
\bottomrule\end{tabular}}\end{table}

\paragraph{The solver domain}
\NumBdShortText\ \NumOptShortText\ The regular-set correction \eqref{eq:regcorr} with $\kappa=\NumVbKappa$, applied post hoc to the stored fits,
differs from the smooth correction by at most \NumRgMaxDiff\ standard errors at the original box, with projection coverage \NumRgCovA, \NumRgCovB\ and
\NumRgCovC. The second study recomputed the wider-box refits with saved fits; its selection rule, the lower criterion of the refit and the original fit,
was fixed after the follow-up, so this is a recomputation of results already seen. With that rule the regular-set correction term
$\widetilde\phi^{\rm reg}_R-\widehat\phi_R$ changed by at most \NumWbTermRegMax\ standard errors (90th percentile \NumWbTermRegQ), against up to
\NumWbTermSmoothMax\ for the smooth term, and projection coverage at $R=256$ was \NumWbCovWide\ of 120 on both domains. The joint estimate still moved by
more than half a standard error in \NumWbJointBig\ of the \NumWbN\ data sets (at most \NumWbJointMax), to other local minima along the slow-input valley
(Supplement~S20). The instability of the correction came from channels near degenerate faces, which the regular set excludes; the
multiplicity of local minima remains.

\paragraph{Second validation study}
After the first study we fixed the computed estimator: the box $\log\beta\in[-6,3]$, the diagnostic input chart $\Phi_C$ (delay within the onset frame less
$1/4$\,s at each end, $|\lambda_1|\ge1.5|\lambda_2|\ge1.5^2|\lambda_3|$, other coordinates within $\pm3$ of the reference input), the regular-set
correction with $\kappa=\NumVbKappa$, and the original Gauss--Newton (primary) and profile-Hessian variance calculations. The threshold is below the empirical minimum \NumValAtA\ from 5000 draws; this is not a certified bound on the support or a verification that $\kappa<c_0$. It was applied to new prespecified data sets at $R=256$, where the first study fell shortest, with $\sigma=0.2$, 0.1 and 0.05
and 400, 200 and 200 independent data sets, so that a 95\% coverage has Monte Carlo standard error 1.1--1.5 points; the design file and code
fingerprints were written before the first run (Supplement~S20). All \NumVbN\ data sets were computed and \NumVbFailed\ failed. No
fit left the chart. The solver checks this chart at the returned fit but does not enforce it along the search or certify minimization over it. Irregular channels occurred only at $\sigma=0.2$ (in \NumVbIrrA\ of 400 data sets,
\NumVbBoxA\ of them with a channel at the box), where the regular-set and smooth corrections differed by at most \NumVbRegSmoothMax\ standard errors.

At $\sigma=0.05$ the observed corrected coverage was compatible with 95\%, \NumVbCovC\ for the projection (exact interval \NumVbCovCCI) and \NumVbCoordC\ for
the six coordinates, with spread ratios \NumVbSdC; the joint intervals covered the projection in \NumVbCovJC. At $\sigma=0.1$ corrected coverage was
\NumVbCovB\ (\NumVbCovBCI) for the projection and \NumVbCoordB\ coordinatewise, with spread ratios \NumVbSdB. At $\sigma=0.2$ the shortfall of the first study reappeared: projection
\NumVbCovA\ (\NumVbCovACI), coordinates \NumVbCoordA, spread ratios \NumVbSdA, and a residual bias of \NumVbBiasA\ standard errors. A post hoc audit replaced the full-tangent projected scores by the envelope scores $-B_r^Te_r$ in the fixed-face profile-Hessian calculation (Supplement~S23). This gave coverage \NumVbCovHA\ for the projection and \NumVbCoordHA\
coordinatewise, with spread ratios \NumVbSdHA; at the two smaller noise levels it changed little (Table~\ref{tab:validation2}). These results support a smaller variance discrepancy as $\sigma$ decreases at fixed $R$. They attribute part of the discrepancy to the variance approximation, without establishing finite-noise validity at active faces; the residual bias at $\sigma=0.2$ is not removed by either sandwich. A second start changed the joint estimate
by more than 0.1 standard errors in \NumVbMultiFar\ of \NumVbMultiN\ data sets, by at most \NumVbMultiMax, and the standardized score exceeded 0.1 in
\NumVbScoreOver\ data sets (maximum \NumVbScoreMax), all at $\sigma=0.2$.

\begin{table}[tb]
\caption{Second validation study, $R=256$: the computational procedure fixed after the first study (box $\log\beta\in[-6,3]$, input chart checked at returned fits,
regular-set correction \eqref{eq:regcorr} with $\kappa=0.01$) on new data sets prespecified before the runs. Sets: number of data sets. All coverages are percentages of nominal 95\% intervals. Projection on the
predicted direction: coverage of joint-estimator intervals; mean $c/\sigma^2$ of the corrected estimator (Monte Carlo standard error)
and its bias over the mean standard error. Corrected intervals with the original Gauss--Newton calculation and the post hoc envelope-score profile-Hessian calculation: projection coverage [exact 95\% interval], range of coordinatewise coverage, and the ratio of empirical standard deviation to mean standard
error for the projection. The post hoc variant changes only the covariance, not the point estimates; the original secondary results are retained in the Supplement.}
\label{tab:validation2}\centering\footnotesize\setlength{\tabcolsep}{3pt}
\begin{tabular}{rrccccccccc}\toprule
 & & joint & \multicolumn{2}{c}{corrected} & \multicolumn{3}{c}{Gauss--Newton sandwich} & \multicolumn{3}{c}{profile Hessian (post hoc)}\\
\cmidrule(lr){4-5}\cmidrule(lr){6-8}\cmidrule(lr){9-11}
$\sigma$ & sets & cov. & $c/\sigma^2$ & bias/se & cov.\ [CI] & coord. & sd/se & cov.\ [CI] & coord. & sd/se\\\midrule
0.2 & 400 & 35 & 0.35 (0.03) & 0.69 & 85 [81, 88] & 72--95 & 1.26 & 90 [87, 93] & 89--96 & 1.11\\
0.1 & 200 & 79 & 0.20 (0.07) & 0.21 & 93 [88, 96] & 91--97 & 1.06 & 93 [89, 96] & 94--98 & 1.03\\
0.05 & 200 & 92 & 0.06 (0.14) & 0.03 & 94 [89, 96] & 93--96 & 1.02 & 94 [89, 96] & 94--97 & 1.01\\
\bottomrule\end{tabular}\end{table}

\paragraph{Heterogeneous channels}
Two post hoc algebraic approximations illustrate why pooling and inversion cannot be interchanged: the average
$-\frac12R^{-1}\sum_rN_r^+m_r$ and the coefficient at the mean channel parameters. At the true input and $2\times10^4$ compact-law channels their
projections on the pooled direction are \NumHetCavg\ and \NumHetCrep\ (pooled value 1), at angles \NumHetAngAvg$^\circ$ and \NumHetAngRep$^\circ$
in the $\bar N$ metric. The first approximation is numerically fragile because individual information matrices have median condition number
\NumHetCond. Supplement~S21 reports their plug-in interval summaries. Neither approximation implements a published template estimator;
these comparisons do not establish superiority over the methods of \citet{miolane2017}. The well-conditioned example in Section~\ref{sec:bias}
provides an independent algebraic demonstration of the weighting issue.

\paragraph{Numerical error}
In the first study, refits from the true input agreed with the fits from the image-derived start within 0.1 standard errors in \NumValOracleNclose\
cases and otherwise ended in other local minima at most \NumValOracleMax\ standard errors apart; a twice finer grid changed no estimate by more than
\NumValFineMax\ standard errors (Supplement~S19).

\paragraph{Physical constraints in the joint fit}
A random quarter of the reference-design channels was given no trapping ($k_3=0$, hence $a_1=0$). On matched data, a fit imposing $a_1\ge0$ was
compared with the smooth macro fit, with $R=2048$ and \NumBdReps\ data sets per noise level (Table~\ref{tab:boundary}). The constrained-minus-unconstrained
difference divided by $\sigma$ tracks the root of Theorem~\ref{thm:boundary} at smaller noise. For $\log|\lambda_3|$ at $\sigma=\NumBdSigLo$, the
measured value is \NumBdLamMeas\ against \NumBdLamPred\ predicted, whereas the curvature contribution has size $|\sigma b|=\NumBdLamCurv$.

\begin{table}[tb]
\caption{Physical constraints in the joint fit (Theorem~\ref{thm:boundary}): a quarter of the XCAT channels have $k_3=0$, $R=2048$. Paired
difference between the joint estimates with and without the constraint $a_1\ge0$, divided by $\sigma$ (mean over data sets, Monte Carlo
standard error), and the prediction $\delta_\sigma$ of Theorem~\ref{thm:boundary}. The last column is the unconstrained curvature bias per unit
$\sigma^2$ in this design, for comparison. The delay and $\log|\lambda_1|$ rows are below $0.02$ in all cells.}
\label{tab:boundary}
\centering\footnotesize\setlength{\tabcolsep}{4pt}
\begin{tabular}{lrrrrrrr}
\toprule
 & \multicolumn{2}{c}{$\sigma=0.025$} & \multicolumn{2}{c}{$\sigma=0.05$} & \multicolumn{2}{c}{$\sigma=0.1$} & curvature\\
\cmidrule(lr){2-3}\cmidrule(lr){4-5}\cmidrule(lr){6-7}
coordinate & measured & predicted & measured & predicted & measured & predicted & $b$\\
\midrule
$\log A_2$ & $0.124$ (0.003) & $0.122$ & $0.121$ (0.004) & $0.122$ & $0.128$ (0.008) & $0.122$ & $-1.56$\\
$\log A_3$ & $-0.072$ (0.003) & $-0.065$ & $-0.067$ (0.005) & $-0.065$ & $-0.065$ (0.005) & $-0.065$ & $-3.07$\\
$\log|\lambda_2|$ & $0.039$ (0.003) & $0.034$ & $0.039$ (0.004) & $0.034$ & $0.056$ (0.004) & $0.034$ & $1.42$\\
$\log|\lambda_3|$ & $1.712$ (0.053) & $1.626$ & $1.604$ (0.043) & $1.626$ & $1.793$ (0.101) & $1.626$ & $-2.96$\\
\bottomrule
\end{tabular}
\end{table}

\paragraph{Population models and stress tests}
Local quadrature diagnostics for working Gaussian mixtures agree with the population-score coefficient within \NumPopRefSmall\ at $\sigma=0.01$
(Supplement~S22). Under misspecification (reversible kinetics, channel-specific delays or dispersion, a varying plasma-to-blood
ratio) the noise-free limit of the joint estimator moved by more than the curvature bias at $\sigma=0.2$ in every case but the last (Supplement,
Table~S28).

\paragraph{Correlated noise and covariance calibration}
For the flat channels of Example~\ref{ex:flatcov}, known covariance separates the roles of bias correction and spatial variance adjustment (Supplement~S24). With $R=1024$ and marginal and spatial correlations both $0.6$, corrected 95\% coverage at $\sigma=0.2$, 0.1 and 0.05 was 94.27, 94.88 and 95.06\%, against 71.23, 72.29 and 72.53\% when spatial correlation was ignored; correct marginal GLS gave similar coverage. Estimating the covariance from one independent calibration field, as in Proposition~\ref{prop:covcalibration}, gave \NumCalFlatCoverage\% (10,000 replicates per cell). The correction repairs the specified Euclidean estimator; no superiority over correct whitening is claimed.

A compartment-model experiment uses \NumCalPetEach\ independent datasets at each of $\sigma=0.1$ and 0.05, with $R=256$, temporal correlation $0.5$, spatial correlation $0.4$ and one independent calibration field per dataset; all datasets are retained. The estimated temporal matrix had a mean relative operator-norm error of about \NumCalPetCovRelError\%, yet within each dataset the estimated-covariance and oracle analyses agreed closely on the standard-error scale: ratios of their standard errors had median \NumCalPetSeRatioMed\ and 5--95\% range \NumCalPetSeRatio, and the corrected contrasts differed by at most \NumCalPetErrDiff\ oracle standard errors. Coverage of the prespecified contrast was \NumCalPetRawCov\% without correction, \NumCalPetOracleCov\% with oracle correction and \NumCalPetEstimatedCov\% with estimated correction. The oracle and estimated intervals disagreed for \NumCalPetDiscordant\ of 50 datasets at $\sigma=0.1$ and 0.05, respectively, all borderline, so the study does not rank the two analyses; with 50 replicates per level its coverage estimates are imprecise (Supplement~S24.4). These simulations assume known fitting weights and a correctly specified calibration field, neither of which is validated for a patient reconstruction.

\section{Discussion}\label{sec:discussion}
The theoretical distinction is between identification and estimation with incidental channels. A shared input can be identifiable from frame means
yet estimated with a persistent curvature bias, whose leading term is an information-weighted regression of mean curvature on normal input velocity.
The analytic correction removes this term under uniform smoothness and small-noise conditions; it is not a fixed-noise consistency result. Boundary
constraints and population misspecification introduce different bias terms. With general marginal covariance, even flat moving channels can be biased. The Gaussian-dependence theorem distinguishes this marginal effect from the full score covariance needed for uncertainty.

In compact-population PET simulations with prespecified independent replicates, the joint intervals deteriorated as $R\sigma^2$ grew and the
correction removed most of the bias. The two validation studies locate where the first-order theory is accurate. Along $\sigma_R\propto R^{-1/4}$,
the corrected intervals retained a residual bias of 0.5--0.7 standard errors and undercovered at $R=256$ and 1024. At $R=256$ with the estimator
fixed in advance, they were close to nominal at $\sigma=0.05$ and 0.1 but not at $\sigma=0.2$, where a residual bias remained and the Gauss--Newton
sandwich understated the variance; the audited fixed-face profile-Hessian calculation reduced that understatement. Its score correction was post hoc, and it supplies no boundary-limit theorem. The finite-design expansion of the
Supplement accounts for much of the gap between the fitted estimator and the population coefficient at fixed $R$. We therefore claim calibrated
inference only in the small-noise regime of Theorem~\ref{thm:corrected}; at $\sigma=0.2$ in this design the corrected intervals are better than the
uncorrected ones but not calibrated.

The computed estimator includes its domain. A box that lets fitted channels approach degenerate faces of the channel family makes the smooth plug-in
correction unstable; the regular-set correction, equivalent to it under the assumptions, substantially reduced that instability of the correction term in the tested fits. Choosing
the lower criterion among candidate fits keeps the reported optimum from worsening when the domain grows, although the joint fit can then move to
another local minimum. Weak identification of the slow input components still makes the
profiled criterion nearly flat along a valley, so starting values and optimization diagnostics matter at moderate noise. \NumOptDiscText\ Finally, heterogeneity makes the order of pooling and inversion important: the two-type example proves this without ill-conditioning, and the PET diagnostics illustrate its numerical consequences.

Unrestricted non-Gaussian spatial dependence, estimation of reconstruction covariance without external information, multiple active constraints, a uniform bootstrap theory and reliable estimation of flexible populations remain outside the results; the hierarchical sampler we implemented did not converge. The historical single-subject PET example (Supplement~S30), computed
with the earlier kernel, has unresolved optimization, correlated reconstructed noise and uncertain calibration; it illustrates sensitivity and supports
no calibrated-inference claim.

\begin{acks}[Acknowledgments]
The author thanks the developers of PET-TRACER, vPET-ABC, \texttt{dynamicpet} and \texttt{qurit/dynamicPET} for making their code
and data openly available.
\end{acks}

\begin{funding}
H.-H. Huang acknowledges support from the National Science Foundation under grants
DMS-1924792 and DMS-2318925 and from the National Institutes of Health (NIH), National
Institute of Allergy and Infectious Diseases (NIAID) R01 award (1R01AI206603-01).
\end{funding}

\begin{supplement}
\stitle{Supplement to ``Identification and curvature bias for shared inputs in multichannel dynamical models''}
\sdescription{Available from the author on request: proofs, additional theoretical results, simulation details and numerical diagnostics, the historical PET sensitivity analysis, and reproducibility code and data-access information.}
\end{supplement}

\bibliographystyle{spr-ims-nameyear}
\bibliography{references}
\end{document}